# Numerical simulations of surf zone wave dynamics using Smoothed Particle Hydrodynamics

R. J. Lowe[1,2,3,4], M. L. Buckley[1,4], C. Altomare[5,6], D. P. Rijnsdorp[1,4], Y. Yao[7], T. Suzuki[8,9], J. Bricker[9]

[1] Oceans Graduate School and UWA Oceans Institute, The University of Western Australia, Crawley, Australia
[2] School of Earth Sciences, The University of Western Australia, Crawley, Australia
[3] ARC Centre of Excellence for Coral Reef Studies, The University of Western Australia, Crawley, Australia
[4] Wave Energy Research Centre, The University of Western Australia, Crawley, Australia
[5] Universitat Politècnica de Catalunya – Barcelona Tech, Barcelona, Spain
[6] Ghent University, Ghent, Belgium
[7] Changsha University of Science and Technology, Changsha, China
[8] Flanders Hydraulics Research, Antwerp, Belgium
[9] Dept. of Hydraulic Engineering, Faculty of Civil Engineering and Geosciences, Delft University of Technology, Delft, Netherlands

Corresponding author:

Ryan J. Lowe (Ryan.Lowe@uwa.edu.au)
The University of Western Australia
35 Stirling Hwy, Crawley, Western Australia, Australia

**Abstract**

In this study we investigated the capabilities of the mesh-free, Lagrangian particle method (Smoothed Particle Hydrodynamics, SPH) to simulate the detailed hydrodynamic processes generated by both spilling and plunging breaking waves within the surf zone. The weakly-compressible SPH code DualSPHysics was applied to simulate wave breaking over two distinct bathymetric profiles (a plane beach and fringing reef) and compared to experimental flume measurements of waves, flows, and mean water levels. Despite the simulations spanning very different wave breaking conditions (including an extreme case with violently plunging waves on an effectively dry reef slope), the model was able to reproduce a wide range of relevant surf zone hydrodynamic processes using a fixed set of numerical parameters. This included accurate predictions of the nonlinear evolution of wave shapes (e.g., asymmetry and skewness properties), rates of wave dissipation within the surf zone, and wave setup distributions. By using this mesh-free approach, the model is able to resolve the critical crest region within the breaking waves, which provided robust predictions of the wave-induced mass fluxes within the surf zone responsible for undertow. Within this breaking crest region, the model results capture how the potential energy of the organized wave motion is initially converted to kinetic energy and then dissipated, which reproduces the distribution of wave forces responsible for wave setup generation across the surf zone. Overall, the results reveal how the mesh-free SPH approach can accurately reproduce the detailed wave breaking processes with comparable skill to state-of-the-art mesh-based Computational Fluid Dynamic models, and thus can be applied to provide valuable new physical insight into surf zone dynamics.

## 1  Introduction

The accurate prediction of wave transformation and wave breaking in the nearshore zone, including how waves impact coastal structures, remains one of the great challenges in the fields of nearshore oceanography and coastal engineering. Much of this uncertainty stems from how to most accurately simulate the breaking process (i.e., overturning of the free surface), and in turn, how wave transformation within the surf zone region triggers a range of additional nearshore hydrodynamic processes, including additional sources of water level variations (e.g., low frequency waves and wave setup) and wave-driven mean flows. The accurate description of the full range of nearshore water motions is critical to develop robust predictions of wave-driven coastal impacts, including coastal flooding, erosion and storm damage to coastal infrastructure.

The nonlinear physics that govern nearshore wave transformation (e.g., the cross-shore evolution of wave shape, nonlinear energy transfers, and ultimately dissipation) are especially challenging to predict in practical coastal-scale applications due to the wide range of spatial and temporal scales of the processes involved. For example, both breaking and non-breaking waves drive mass transport over relatively large-scales (i.e., order 10s to 100s of meters); whereas, incident wave energy ultimately becomes dissipated as heat within turbulent flow fields at much finer-scales (i.e., order centimetres or less). Historically, it has been impractical to directly predict (both analytically and numerically) the full spectrum of hydrodynamic processes in the nearshore zone. To fill this gap, experimental observations (both within the lab and field) have been critical to advance process understanding of nearshore hydrodynamics by supporting the development of empirical formulations to parameterize surf zone processes that occur at scales finer than can be resolved by a coastal model. However, as a general rule, a reliance on the parameterization of physical processes within coastal models can risk undermining their predictive benefits; for example, this can require case-specific (non-physical) tuning of empirical parameters to datasets or may cause models to entirely fail when extended to coastal applications and/or or complex study sites beyond the parameter space for which they were initially developed and validated.

Nearshore wave models can be broadly placed into two main categories: phase-averaged (spectral) and phase-resolving models. Phase-averaged wave models attempt to simulate the stochastic properties of waves, usually based on linear wave theory, with empirical formulations to parameterize the nonlinear physics (e.g., wave breaking dissipation, wave-wave interactions, etc.). These models are also commonly coupled to flow models to simulate slowly-varying flow properties (i.e., time-scales greater than the wave group envelope period) such as wave setup and mean wave-driven currents. Given that phase-averaged models can only provide crude representations of complex surf zone physics, they often require some degree of parameter tuning to match experimental observations and thus may be incapable of simulating the full range of hydrodynamic processes that are important in a nearshore application. For example, it is relatively common to find that models that have been tuned to optimally reproduce surf zone wave transformation underpredict the magnitude of wave setup, especially when waves break on steep slopes (e.g., Lashley et al., 2018, Skotner and Apelt, 1999). This indicates that these models may not correctly predict the cross-shore distribution of wave forces (radiation stress gradients), which has been attributed to the poor accounting of the conversion of potential to kinetic energy under breaking waves and subsequent dissipation in the inner surf zone (e.g., Buckley et al., 2015). The inclusion of additional empirical formulations (e.g., roller models) have been proposed to account for these unresolved processes during breaking (Svendsen, 1984); however, these approaches have had varying degrees of success and introduce additional model parameters that are not universally applicable across nearshore applications.

Phase-resolving wave models, loosely defined here as any wave-flow model that deterministically resolve motions at time-scales shorter than individual sea-swell waves, are designed to include a more complete representation of the nonlinear physics of nearshore waves. Depth-averaged (2DH, two-dimensional in the horizontal) versions of these models (or multi-layer versions employed with coarse vertical resolution) are commonly either Boussinesq-type or based on the nonlinear shallow water equations with non-hydrostatic pressure corrections (commonly referred to as non-hydrostatic wave-flow models). While 2DH phase-resolving models may more accurately simulate the nonlinear behaviour of non-breaking waves in shallow water, they are still incapable of directly resolving the wave breaking process and thus may suffer from some of the same shortcomings as phase-averaged models (i.e., empirical parameterization of the breaking process). Due to their (quasi) depth-averaged description, 2DH models also do not capture the vertical structure of the (mean) flow dynamics (e.g., the undertow profile). Three-dimensional (3D) phase-resolving wave-flow models (summarized below) provide the most rigorous representation of nearshore hydrodynamics, including the capability to directly resolve at least some of the wave breaking process, but are computationally expensive. These models can be classified as either mesh-based or mesh-free. The former class are based on various solutions of Eulerian forms of the (Reynolds-Averaged) Navier-Stokes equations on numerical grids (meshes). The latter class are based on Lagrangian solution of the Navier-Stokes equations, which include those based on particle methods that attempt to describe the motion of a fluid continuum using discrete particles (Liu and Liu, 2003).

Several 3D mesh-based approaches have been developed to simulate waves in the nearshore, which mainly differ in their treatment of the free surface. Multi-layered non-hydrostatic wave-flow models (e.g., Zijlema and Stelling, 2008, Bradford, 2010, Ma et al., 2012) describe the free-surface as a single-valued free-surface (akin to 2DH phase-resolving models). Although this simplification allows them to capture the dissipation of breaking waves and 3D flows more efficiently, it also implies that they cannot resolve all details of the breaking process such as wave overturning, and breaking wave-generated turbulence (e.g., Derakhti et al., 2016b, Rijnsdorp et al., 2017). Alternatively, Computational Fluid Dynamic (CFD) models use more comprehensive techniques that can capture complex details of the free surface such as under breaking waves. This includes the marker and cell method (Harlow and Welch, 1965), level-set method (Osher and Sethian, 1988), and Volume Of Fluid (VOF) method (Hirt and Nichols, 1981), of which the VOF approach has been most widely adopted to model nearshore processes (e.g., Torres‐Freyermuth et al., 2007, Yao et al., 2019). Although these models can capture the breaking process and the turbulent flow field in detail, comparisons between laboratory experiments and mesh-based RANS solvers have highlighted some of the difficulties in accurately predicting turbulent flow fields within the surf zone. In particular, model predictions have been shown to be sensitive to the turbulence model used (e.g., Brown et al., 2016), which can lead to a significant discrepancies in undertow profiles predicted throughout the surf zone (e.g., Brown et al., 2016, Larsen and Fuhrman, 2018).

The most common mesh-free particle methods are Smoothed Particle Hydrodynamics (SPH) and Moving Particle Semi-implicit (MPS) models. SPH methods (which is the specific focus here), were originally developed as a general numerical approach for supporting continuum mechanics applications, which are now used across a range of scientific fields, including astrophysics, fluid mechanics and solid mechanics (Monaghan, 1992). In more recent years (particularly over the past decade), SPH has become an increasingly common technique applied to coastal and ocean engineering problems, due to its ability to deal with complex geometries, account for highly nonlinear flow behaviour, and to simulate large deformations at interfaces (including moving boundaries and at the free surface) (e.g., Monaghan and Kos, 1999, St-Germain et al., 2013, Altomare et al., 2015b, Crespo et al., 2017, González-Cao et al., 2019, Zhang et al., 2018a, Altomare et al., 2014, Domínguez et al., 2019). Nevertheless, SPH is still being continuously developed and improved; for example, there is still concerted international effort to improve performance related to numerical implementations (including enhancing model convergence, consistency, and stability and adaptivity schemes) and improving treatment of boundary conditions (including at solid boundaries, at the free surface and when coupling to other models). In coastal wave applications, SPH techniques are now being increasingly applied to study wave-structure interactions (including loads and overtopping) (e.g., Akbari, 2017, Altomare et al., 2015a, González-Cao et al., 2019) and the dynamics of floating bodies (e.g., Bouscasse et al., 2013, Ren et al., 2017, Crespo et al., 2017). To a lesser degree, SPH approaches have also been used to investigate the physics of surf zone processes; for example, in recent studies of surf zone currents and eddies (Wei et al., 2017, Farahani et al., 2013), nearshore wave breaking (e.g., Issa and Violeau, 2009, Makris et al., 2016, De Padova et al., 2018, Roselli et al., 2019, Shao and Ji, 2006) and surf zone energy balances (Wei and Dalrymple, 2018). Nevertheless, despite the great promise of SPH to nearshore applications, a rigorous assessment of the ability of SPH models to accurately simulate a full range of relevant surf zone hydrodynamic processes is still relatively sparse, certainly in comparison to the wealth of information derived from detailed nearshore wave modelling studies (including mesh-based CFD modelling approaches).

In this study, we conduct a detailed investigation of the ability of the SPH modelling approach to predict a broad range of nearshore processes relevant to coastal applications where wave breaking is important. Using experimental data of wave breaking over both a plane beach and a fringing reef profile, we demonstrate how the model can accurately reproduce a broad range of relevant hydrodynamic processes, ranging from the nonlinear evolution of wave shapes across the surf zone, wave setup distributions, and mean current profiles. We compare the present surf zone predictions with predictions by other classes of wave models from literature and illustrate some of the advantages of the SPH approach (particularly in resolving hydrodynamics within the crest region above the wave trough).

## 2  Methods

### 2.1 General features of the SPH method

Smoothed Particle Hydrodynamics (SPH) is a mesh-free numerical method where a continuum is discretised into particles. The approach was originally developed within astrophysics (Lucy, 1977, Gingold and Monaghan, 1977) and since then largely applied across a wide range of Computational Fluid Dynamics (CFD) applications. Within SPH, the particles represent calculation nodal points that are free to move in space according to the governing Lagrangian dynamics, such as in fluid mechanics based on the Navier-Stokes equations (Monaghan, 1992). The kinematic and dynamic properties of each particle (e.g., position, velocity, density pressure, etc.) then result from the interpolation of the values of the neighbouring particles. The distance between each *i-th* particle and its neighbours determines the weighting of the contribution of the nearest particles based on application of a weighted kernel function (*W*). The area of influence of the kernel function is defined using a characteristic smoothing length ($h_{SPH}$). The kernel function is a finite representation of the Dirac function (i.e., for the limit where $h_{SPH}$ approaches zero), and has a finite distance cut-off (often $\pm 2$ or 3 times $h_{SPH}$) to avoid contributions (and hence interaction computations) with other particles beyond this distance. While a variety of kernel functions have been proposed for SPH, a Quintic kernel (Wendland, 1995) is often used (including in the present study), where the weighting function vanishes for initial particle spacing greater than $2h_{SPH}$.

In general, SPH methods can be grouped into two main classes: Weakly Compressible SPH (WCSPH) and Incompressible SPH (ISPH). A comprehensive review of both the WCSPH and ISPH approaches is presented in Gotoh and Khayyer (2018), which describes the latest developments of both WCSPH and ISPH in terms of stability, accuracy, energy conservation, boundary conditions and simulations of multiphase flows and fluid–structure interactions. The fundamental difference between WCSPH and ISPH is how each method solves for the pressure and density fields. In WCSPH, an appropriate equation of state (i.e., Tait's equation, see below) is solved in a fully-explicit form; whereas Incompressible SPH (ISPH) solves a Poisson pressure equation by applying projection-based methods. The primary advantage of WCSPH is the ability to directly relate pressure and density, rather than having to obtain pressure fields by solving a Poisson equation (as in the case of ISPH) at significant computational expense. As a consequence, WCSPH can be readily parallelized in numerical codes, including on Graphics Processing Units (GPUs) (see below), given that the motion of each particle is solved independently. The WCSPH approach, however, is not without drawbacks (Lee et al., 2008): it can require using very small numerical time steps, and also require using various numerical approaches to avoid model instabilities due to non-physical density / pressure fluctuations that can arise from the compressibility of the flow (see below). For the present work we use a WCSPH solver, so the focus will be on detailing specific aspects of this approach further below.

The mathematical foundation of SPH is based around integral interpolants, in which any function $F(\vec{r})$ in coordinate space $\vec{r}$ can be computed by the integral approximation:

$$F(\vec{r}) = \int F(\vec{r}')W(\vec{r}-\vec{r}',h_{SPH})d\vec{r}' \qquad (1)$$

This function $F$ can be expressed in discrete form based on particles, in which the approximation of the function is interpolated at particle $a$ and the summation is performed over all the particles $b$ that are located within the region of the kernel:

$$F(\vec{r}_a) \approx \sum_b F(\vec{r}_b)W(\vec{r}_a-\vec{r}_b,h_{SPH})\Delta V_b \qquad (2)$$

where the volume $\Delta V_b$ associated with a neighbouring particle $b$ is $m_b/\rho_b$, where $m$ and $\rho$ are mass and density, respectively. The momentum equation in discrete SPH form for a weakly compressible fluid can then be written as (Monaghan, 1992):

$$\frac{d\vec{u}_a}{dt} = -\sum_b m_b \left( \frac{P_b+P_a}{\rho_b \rho_a} + \Pi_{ab} \right) \nabla_a W_{ab} + \vec{g} \qquad (3)$$

where $t$ is time, $\vec{u}_a = d\vec{r}_a/dt$ is the velocity of particle $a$, $P$ is pressure, $\vec{g}$ is gravitational acceleration, and $W_{ab}$ is the kernel function that depends on the distance between particles $a$ and $b$. The effect of viscous dissipation within SPH can be approximated using the artificial viscosity term $\Pi_{ab}$ (Monaghan, 1992):

$$\Pi_{ab} = \begin{cases} \dfrac{-\alpha \overline{c_{ab}} \mu_{ab}}{\overline{\rho_{ab}}} & \vec{u}_{ab} \cdot \vec{r}_{ab} < 0 \\ 0 & \vec{u}_{ab} \cdot \vec{r}_{ab} > 0 \end{cases} \quad (4)$$

where $\vec{r}_{ab} = \vec{r}_a - \vec{r}_a$ and $\vec{u}_{ab} = \vec{u}_a - \vec{u}_b$, $\mu_{ab} = h_{SPH} \vec{u}_{ab} \cdot \vec{r}_{ab} / (\vec{r}_{ab}^2 + \varepsilon^2)$ with $\varepsilon^2 = 0.01 h_{SPH}$, $\overline{c_{ab}} = 0.5(c_a + c_b)$ is the mean speed of sound, and $\alpha$ is a coefficient (termed artificial viscosity) that determines the rate of viscous dissipation. The formulation for $\Pi_{ab}$ given by Eq. (4) is linearly proportional to velocity gradients and thus produces the effect of a shear and bulk viscosity. Altomare et al. (2015a) proposed using a reference value of $\alpha = 0.01$ for coastal applications, based on model validation against experimental data for wave propagation and induced loading onto coastal structures. Roselli et al. (2018) employed a Multi-Objective Genetic Algorithm to find a set of SPH parameters that led to accurate modelling of wave propagation and found a comparable optimal value for $\alpha$ equal to 0.004. While more sophisticated formulations have been proposed to predict viscous stresses, including Sub-Particle Scale (SPS) turbulence closure models (Gotoh et al., 2004, Dalrymple and Rogers, 2006) that are analogous to Large Eddy Simulation (LES) in CFD applications, in the present study we use the simple artificial viscosity scheme with high particle resolution, which we found led to more robust (stable) model simulations over the range of test cases considered (see section 2.4) with accurate results.

In a similar way to Eq. (3), a discrete form of the continuity equation can be expressed as:

$$\frac{d\rho_a}{dt} = \sum_b m_b (\vec{u}_a - \vec{u}_b) \cdot \nabla_a W_{ab} \quad (5)$$

For WCSPH, following Monaghan (1994), pressure and density are related by Tait's equation of state that is expressed as:

$$P = B\left[\left(\frac{\rho}{\rho_0}\right)^\gamma - 1\right] \quad (6)$$

where $\gamma = 7$ is the polytropic constant and $B = c_0^2 \rho_0 / \gamma$ is defined based on the reference density $\rho_0$ and the speed of sound $c_0$. The equation of state is considered stiff, so oscillations in the density field are allowed within a range of 1% by adjusting the compressibility (and hence the speed of sound) and finding a trade-off between the size of the time step (determined by the Courant condition based on the speed of sound) and the density variations. This is achieved by reducing the speed of sound but maintaining it at least 10 times higher than the maximum velocity in the system to approximate an incompressible flow. Note that while an additional diffusive term is often included in Eq. (5) to reduce density fluctuations, such as through incorporation of a $\delta$-SPH formulation (Molteni and Colagrossi, 2009), such approaches can lead to some detachment of the free surface due to truncation of the kernel close to the free surface within breaking waves; therefore, to avoid introducing potential inaccuracies, in the present study we did not use the $\delta$-SPH formulation within the version of DualSPHysics used.

Finally, in the SPH framework any number of derived hydrodynamic quantities can be defined and computed. For example, the vorticity $\vec{\omega} = \nabla \times \vec{u}$ (examined later in this study) can be computed for an arbitrary particle $i$ based on gradients computed with surrounding particles $j$ within the compact support domain (Monaghan, 1992):

$$\omega_i = \sum_j m_j \left( \vec{u}_i - \vec{u}_j \right) \times \nabla_i W_{ij}. \tag{7}$$

**2.2 The DualSPHysics model**

The present study uses the open-source solver DualSPHysics (http://dual.sphysics.org/) based on WCSPH (Crespo et al., 2011, Crespo et al., 2015). DualSPHysics is written in two languages, namely C++ and CUDA, and optimized to use the parallel processing power of either CPUs and/or GPUs (Domínguez et al., 2013). GPUs offer greater computing power relative to CPUs, and thus have emerged as an affordable option to accelerate SPH modelling, including making the study of real engineering-scale problems more possible (e.g., Altomare et al., 2014).

The governing mass and momentum equations (Eqs. (3) and (5)) were numerically solved within DualSPHysics by integrating in time using a numerically stable two-stage explicit Symplectic method with a variable time step that was functionally dependent on a combination of the Courant–Friedrich–Levy (CFL) condition, the forcing terms and the viscous diffusion term following (Monaghan and Kos, 1999) (see Crespo et al., 2015 for details of the specific implementation). Within SPH, a single (optimum) approach to incorporate solid boundaries in fluid mechanics applications has yet to be established, with improvement of boundary conditions having been specified by the SPH community as a priority research topic, including within the SPHERIC Grand Challenges (http://spheric-sph.org/grand-challenges). In DualSPHysics, the interaction between solid boundaries and fluid particles is solved by employing Dynamic Boundary Conditions (DBCs), as described in Crespo et al. (2007), which provides a very simple yet robust method for incorporating fluid-solid interactions that can be easy to apply, even for very complex geometries. With this approach, solid boundaries consist of a set of particles that are treated as fluid particles but their movement is constrained: the boundaries can be fixed or can move according to a particular forcing or motion time series. The Navier-Stokes equations are then applied to the interactions between boundary and fluid particles, with the only exception that the movement of the boundary particles is prescribed or equal to zero. When a fluid particle approaches a boundary particle, the fluid density locally increases, which in turn generates an increase in the pressure per the equation of state (Eq. (6)) employed in WCSPH. The rise of pressure results in a repulsive force that prevents a fluid particle from passing through the boundary particles. In DualSPHysics the user has the option to apply additional artificial viscosity to the fluid-boundary interaction, which can be used to parameterize drag forces at the boundary (i.e., due to sub-particle-scale bed roughness). As the experiments considered in this study had smooth walls, no boundary enhancement of artificial viscosity was used in the present study.

For wave generation, DualSPHysics implements different schemes. Waves can be generated using boundary particles as moving boundaries (MB) that mimic the movement of a wavemaker in a physical facility. Long-crested second-order monochromatic waves, random sea states (including bound long waves) and solitary waves can be automatically generated (Domínguez et al., 2019, Altomare et al., 2017). To absorb the reflected waves at the wavemaker and prevent the introduction of extra spurious energy in the fluid domain, an Active Wave Absorption System (AWAS) has been implemented in DualSPHysics (Altomare et al., 2017). The water surface elevation at the wavemaker position is used and transformed by an appropriate time-domain filter to obtain a control signal that corrects the wave paddle displacement in order to absorb the reflected waves. The position of the wavemaker is obtained in real time through the velocity correction of its motion. As an alternative to MB, waves can be generated in DualSPHysics by enforcing the orbital velocity of the fluid particles in a specific generation area, using a Relaxation Zone method described in (Altomare et al., 2018) or imposing fluid velocity and surface elevation in a buffer zone defined within an open boundary scheme (Verbrugghe et al., 2019a). Finally, coupling with other models can also be employed to generate waves in DualSPHysics (e.g., Altomare et al., 2015b, Altomare et al., 2018, Verbrugghe et al., 2019b): for these applications, the water surface elevation and orbital velocity of SPH fluid particles is derived from other phase-resolving wave models (e.g., OceanWave3D and SWASH). In the present study, MBs with a piston-type wavemaker are employed.

### 2.3 Experimental cases

Model performance was assessed using observations from two experimental datasets that include a range of wave conditions and different bathymetry profiles. This study focuses on the hydrodynamics generated by regular (monochromatic) waves, which allows the surf zone processes to be investigated in detail under quasi-steady state conditions with reduced computational times (i.e., simulating order 100 waves versus order 1000 waves to resolve irregular wave statistics). The two experimental datasets comprise (Figure 1): 1) the two test cases of regular wave breaking on a plane beach in Ting and Kirby (1994) (hereafter denoted TK94) that include both 'spilling' and 'plunging' wave breaking conditions; and 2) the two test cases reported in Yao et al. (2012) (hereafter Y12) that report regular wave transformation across a reef profile at two still water depths.

### *Ting and Kirby (1994) – plane beach*

TK94 obtained detailed measurements of the mean and turbulent flow structures generated by regular wave breaking on a linear-sloping plane beach (note that subsequent analysis of this dataset is also described in Ting and Kirby (1995) and Ting and Kirby (1996)). The wave flume was configured with an initial flat bed (depth $h$=0.4 m) followed by a beach with 1:35 slope (Figure 1a). 'Spilling' breaking waves were generated by incident waves with offshore wave height $H$=0.125 m and period $T$=2.0 s; whereas 'plunging' breaking waves were generated by waves of effectively the same height at breaking ($H$=0.128 m) but with longer period ($T$=5 s). For both cases, water elevations ($\eta$) were recorded using capacitance wave gauges at 100 Hz at ~20 cross-shore locations. Vertical profiles of flow velocities (both horizontal $u$ and vertical $w$ velocity components) were obtained using laser Doppler anemometry (LDA) at 100 Hz at several cross-shore locations (8 and 7 locations for the spilling and plunging cases, respectively); note that due to the intermittent wetting / drying and presence of air bubbles during breaking, valid current measurements could not be obtained over the entire crest-to-trough region (only the lower portion). An additional advantage of the TK94 dataset is its historical use to assess the performance of a number of different models, which range from applications using Boussinesq wave models (e.g., Tissier et al., 2012, Cienfuegos et al., 2010), non-hydrostatic RANS models (e.g., Derakhti et al., 2016a, Rijnsdorp et al., 2017, Smit et al., 2013), and mesh-based CFD models such as OpenFOAM (e.g., Jacobsen et al., 2012, Brown et al., 2016), thereby providing an opportunity to place the present SPH simulations in the context of prior studies using other classes of models.

### *Yao et al. (2012) – fringing reef*

Y12 describe measurements of wave transformation across a reef profile with a steep (~1:6) sloping forereef and 7 m wide horizontal reef flat located 0.35 m above the wave flume bottom (Figure 1b). At the back of the reef flat, the depth extended again to the bottom of the flume and was followed by a 1:8 beach covered with a porous mat to dissipate wave energy. Note that in the model configuration we treated this beach as a solid boundary as residual wave energy reaching this location was negligible (see below); i.e. effectively all the incident wave energy was dissipated over the reef. We also confirmed that this downstream configuration had effectively no influence on the results over the reef, given that initial testing of a reflective vertical wall at the location of the beach had no noticeable effect on any of the results measured across the reef.

In this study we focus on two of the test cases described in Y12 (denoted Case 1 and 3 in that study), which used similar regular wave conditions (incident wave heights of $H$=0.095 m and $H$=0.101 m, and periods $T$=1.25 s and 1.00 s for Cases 1 and 3, respectively). The main difference between the cases was the still water level relative to the reef flat: Case 1 had a still-water reef flat depth of $h_r$=0.1 m, whereas for Case 3 the still water level was at the reef flat level (i.e., $h_r$=0 m). For both cases, water levels were measured at 50 Hz using 8 resistance-type wave gauges and 4 ultrasonic water level sensors. In addition, Yao (2012) describes additional data from the same experiment that include (only for Case 1) current profiles measured below wave trough at 13 locations, as well as video imagery of the outer surf zone region near the reef crest recorded at 30 Hz. While the video images were not rigorously georeferenced, several horizontal and vertical reference locations in the images were known, which allowed for a rough transformation of pixel to world coordinates that enabled (at least a qualitative) comparison with the SPH simulations.

**2.4 Model application and numerical settings**

The numerical simulations were conducted using version 4.2 of DualSPHysics in a 2DV (vertical) plane. To simulate the test cases, a numerical wave flume was constructed with a piston-type wave maker located at the offshore boundary (Altomare et al., 2017), which included the exact geometry of the beach in TK94 and reef in Y12. Both experiments were conducted in large wave flumes that included a long uniform-depth region offshore of the beach / reef, thus creating a large volume of water to resolve that greatly increased computational times, despite wave properties being approximately constant across this offshore region. During initial model testing, we found that reducing the length of this offshore regional had a negligible effect on the surf zone hydrodynamics, so we shortened the offshore region by 10 m in the model applications of both experiments.

For the TK94 spilling wave case, the offshore conditions with $H/gT^2 = 0.008$ and $h/gT^2 = 0.03$ fell within the limits of second order wave theory, so second order wave generation was used to drive the piston in DualSPHysics following Altomare et al. (2017). However, the plunging wave case with $H/gT^2 = 0.0005$ and $h/gT^2 = 0.002$ fell within a cnoidal wave regime, so the piston displacement timeseries was prescribed as model input based on cnoidal wavemaker theory following Cho (2003). Both of the Y12 experiments had offshore wave conditions that fell within the approximate limits of second order theory ($H/gT^2 = 0.006-0.010$ and $h/gT^2 = 0.03-0.04$), which was used within DualSPHysics to drive the wavemaker.

To provide a rigorous and transparent assessment of model performance over the range of test cases, we concentrated on using default or typical model parameter settings reported in the literature that were held constant across all numerical results reported here, i.e. we did not vary parameters across different test cases to optimize performance. However, during initial testing we explored the model sensitivity to a subset of parameters (see Appendix A for details), focusing specifically on the effect of: 1) initial inter-particle spacing (*dp*), 2) the artificial viscosity ($\alpha$) and 3) the smoothing length (*h$_{SPH}$*) (see section 2.1).

Prior SPH studies of wave propagation have suggested that the initial inter-particle spacing *(dp)* should be chosen to be at most 1/10 of the wave height or smaller to properly resolve the free surface and hence minimize non-physical wave dissipation (e.g., Roselli et al., 2018). Therefore, based on the incident wave heights (order 0.1 m across all test cases), this indicates that *dp* should be 1 cm or smaller). From out testing, we found minimal improvement in model predictions of wave height, setup and mean currents for *dp*<5 mm (see Appendix A and Supplementary Material); however, for all simulations we conservatively used *dp*=2 mm. This translated to simulating ~1.22 million particles in both TK94 test cases, and ~1.24 million and ~830 thousand particles for Cases 1 and 3 of Y12, respectively.

The smoothing length *h$_{SPH}$* defines a length-scale that governs the size of the kernel that determines particle-particle interactions in SPH (see section 2.1). In 2D, the smoothing length is related to the initial inter-particle spacing according to $h_{SPH} = coefh\sqrt{2}dp$, where *coefh* is a smoothing coefficient of order 1 that determines the scale of interactions with adjacent particles (with typical values in the range of 1.2-1.8; Roselli et al. (2018)). With too low of value of *coefh*, interactions of a given particle with its neighbouring particles diminishes, effectively leading to numerical dissipation that can, for example, lead to some decay in waves propagating within relatively large numerical wave flumes. Conversely, while increasing *coefh* can enhance energy conservation, this increases the kernel size that effectively reduces model resolution by smoothing the results over the length-scale *h$_{SPH}$*. For the present simulations, we found that values of *coefh* between 1.0 – 1.8 had minimal effect on the results; however, wave height and mean current profile predictions were slightly better for smaller values of *coefh* (see Appendix A and Supplementary Material). Thus, in all simulations we used *coefh* = 1.2, which given *dp*=2 mm implies an effective smoothing length of ~3 mm.

The SPH method requires some source of diffusion scheme to limit high frequency noise and model instability. Various formulations have been proposed to introduce viscous dissipation in the Lagrangian momentum equation, including sophisticated attempts to include turbulence closure schemes (i.e., Sub-Particle turbulence models) that is analogous to a Large Eddy Simulation in fixed-mesh models (e.g., Gotoh et al., 2004). Nevertheless, there still remain open questions on how to best parameterize sub-particle scale turbulent motions in SPH (Violeau and Rogers, 2016), and thus it is common to instead use an artificial viscosity (see Eq. (3)) that is designed to simulate a minimal amount of viscosity to keep the numerical scheme stable. In this way, the artificial viscosity coefficient ($\alpha$) should be chosen to be small enough to allow turbulent motions of size $>dp$ to be properly resolved, but large enough to avoid unstable solutions. A wide range of studies have investigated the optimal value of $\alpha$ in numerical wave flume studies and have consistently found this to be of order 0.01 (e.g., De Padova et al., 2014, Roselli et al., 2018, Altomare et al., 2017, Roselli et al., 2019). Through a sensitivity analysis we found that model performance was not significantly influenced by variations in $\alpha$ between 0.005 and 0.02 (see Appendix A and Supplementary Material), and thus chose to use the default value $\alpha=0.01$ recommended in DualSPHysics within all runs.

**2.5 Numerical simulations and post-processing**

Under regular wave conditions, to attain quasi-steady state conditions that are required to be able to resolve wave, mean current and turbulent properties, a number of studies (both experimental and numerical) have found that of order 10 to 100 individual waves should be resolved (e.g., Ting and Kirby, 1994, Jacobsen et al., 2012). For example, TK94 analysed the convergence characteristics of the number ($N$) of waves averaged to accurately resolve both mean currents and turbulence statistics, and found this occurred after roughly $N=40$ wave periods. We thus conservatively ran the SPH simulations for 360 seconds, which corresponded to resolving $N=180$ waves and $N=72$ waves for the TK94 spilling and plunging cases respectively, and $N \approx 300$ waves for the Y12 test cases. Raw properties (e.g., velocities, pressure, etc.) of the field of particles simulated by DualSPHysics were output at 50 Hz.

All numerical simulations were conducted at the Pawsey Supercomputing Centre in Perth, Australia on a supercomputer (Zeus) with GPU-capable nodes. Each compute node on Zeus contains four Intel Xeon E5-2680 v4 2.4GHz (14 core, 28 thread), each with 256 GB RAM and a dedicated Nvidia Tesla P100 GPU (i.e., 4 GPUs per node). As multi-GPU functionality is not currently available in DualSPHysics, to optimise model performance we packed 4 simulations at a time onto each node of Zeus. Under this configuration, the simulation of all four test cases for 360 seconds (real world time) took approximately 63 hr for both the TK94 spilling and plunging cases; and 67 hr and 45 hr for Case 1 and 3 in Y12, respectively.

To compare the simulation results to the fixed (Eulerian) experimental measurements, for an array of points specified on a grid, a Wendland kernel function (averaging length $2h_{SPH}$) was used to interpolate the individual SPH particle properties to the grid (Wendland, 1995). The resolution of the postprocessing grid varied to coincide with instrument measurement locations and/or to investigate specific process-questions using the numerical output. However, to specifically detect free surface positions across flumes to compare with wave height observations, postprocessing with a much finer vertical grid resolution $\Delta z = 1$ mm was used. The vertical free surface positions were detected as the elevation where the averaging kernel was 40% water by mass (a recommended threshold in 2D versus 50% in 3D), which was obtained by identifying the first point that satisfied this criterion when searching upwards in the water column from the bottom. We note that for the case of plunging waves where the free surface overturns, there can be multiple definitions in the specification of a (single valued) "free surface" height (e.g. alternatively the maximum free surface point could be chosen). However, when comparing the results to experimental measurements, it is equally unclear which water level a capacitance / resistance wire wave gauge would precisely record when there is a mixture of water and air along the gauge. Therefore, we chose to adopt the present approach (a common approach incorporated in DualSPHysics), while acknowledging that some discrepancies between the experimental data and model results could be expected within the surf zone of the plunging wave cases.

The raw timeseries of a given property (e.g., velocities, water levels, etc.) were then decomposed into mean and wave (and in the case of velocities, turbulent) contributions by ensemble averaging over the periodic waves (e.g., Nielsen, 1992). Thus, for example, for the horizontal velocity component ($u$), this was decomposed as (e.g., Nadaoka et al., 1989)

$$u(t) = \bar{u} + \tilde{u}(t) + u'(t) \qquad (8)$$

where the overbar (−) denotes averaging over a timescale much greater than the wave period, the tilde (~) denotes the wave (periodic) contribution, and the prime ( ' ) denotes the turbulent ("random") contribution. The wave velocities ($\tilde{u}$) were obtained by subtracting the mean current velocity ($\bar{u}$) from the ensemble-averaged velocities,

$$\tilde{u}_i = \frac{1}{N} \sum_{j=1}^{N} u_{i,j} - \bar{u} \qquad (9)$$

where *i* represents a time (wave phase) counter, *j* a wave series counter and *N* represents the number of waves averaged over. Turbulent velocities are then defined based on removing the mean current and wave contributions from the raw velocity timeseries, $u' = u - \bar{u} - \tilde{u}$.

We note that in the WCSPH approach, density (pressure) fluctuations can introduce some degree of noise that can translate into velocities. To understand the magnitude of this possible noise relative to the turbulent velocities within the surf zone, we conducted a reference simulation with still water (i.e. no wavemaker motion) using the identical geometry, model configuration, and simulation duration that was used in the TK94 runs. Such an approach is analogous to the WCSPH study by Wei et al. (2018), who assessed the background kinetic energy in a still wave flume to investigate the role that small velocity perturbations may play in the chaotic breaking behaviour of waves. For the still water case, we found that the background kinetic energy density was very small (on-average of order $10^{-9}$ m$^2$ s$^{-1}$, not shown), which was approximately six orders of magnitude smaller than typical turbulent kinetic energy densities of order $10^{-3}$ m$^2$ s$^{-1}$ that were predicted within the surf zone (see below). As a consequence, any numerical noise associated with the WCSPH scheme should be negligible relative to the range of turbulent hydrodynamic motions simulated in this study.

Finally, in a similar way to velocity, the timeseries of water elevation ($\eta$) relative to still water level were decomposed into a time-averaged mean water level ($\bar{\eta}$) (i.e., a wave setup contribution) as well as a periodic, phase-dependent wave contribution ($\tilde{\eta}$) by ensemble averaging over the *N* waves. To quantify the cross-shore evolution in the nonlinearity of wave shape, both the skewness (*Sk*) and asymmetry (*As*) were calculated from $\tilde{\eta}$ as (e.g., Ruessink et al., 2012):

$$Sk = \frac{\overline{\tilde{\eta}^3}}{\left(\overline{\tilde{\eta}^2}\right)^{3/2}}, \quad As = \frac{\overline{\tilde{\eta}_H^3}}{\left(\overline{\tilde{\eta}^2}\right)^{3/2}} \qquad (10)$$

where for *As* the subscript *H* denotes the Hilbert transform of $\tilde{\eta}$. The skewness describes how the crest is much more elevated and narrower to the trough; whereas the asymmetry describes how pitched forward/backward (i.e., "saw-toothed") the waves are.

## 3 Results

### 3.1 Wave breaking over a plane beach (Ting and Kirby, 1994)

Although both of the TK94 test cases had breaking wave heights that were approximately the same, the difference in wave period resulted in appreciable differences in how waves broke and transformed within the surf zone (Figure 2). For the "spilling" case, the model predicts a slight overturning of the free surface (albeit a small volume of water) that initiates at a breaking location $x_b \approx 6.5$ m, which evolves into a turbulent bore that propagates shoreward. For the "plunging" case, the waves undergo greater shoaling, the break point is located further inshore ($x_b \approx 7.5$ m) and there is a much larger overturning (hollow) region during breaking. The initial plunging jet then triggers a sequence of splash-up events from the free surface that propagate well into the inner surf zone.

For both test cases, the cross-shore variations in wave heights are accurately predicted, as illustrated here by comparison of the ensemble-averaged crest ($\eta_{max}$) and trough locations ($\eta_{min}$) (Figure 3a,b). Small discrepancies tend to fall within the variability among individual waves, and possibly well-within the uncertainty of the measurements using capacitance wire wave gauges in the aerated surf zone region, which TK94 note may lead to errors that are difficult to quantify. The cross-shore variations in wave setup ($\bar{\eta}$) are also well-predicted by the model (Figure 3c,d). For the spilling case (Figure 3c) there is an abrupt jump in setup at $x \approx 5.5$ m immediately seaward of the breakpoint that is not consistent with the gradual increase in setup through the surf zone that is predicted with the model. As noted in a number of other studies using this TK94 case (e.g., Cienfuegos et al., 2010, Tonelli and Petti, 2010, Smit et al., 2013), a physical explanation for this small jump in setup is unclear and potentially could reflect some measurement artefact for the wave gauge(s) near the breakpoint.

Figure 4 shows the wave water level timeseries ($\tilde{\eta}$), which are ensemble-averaged over a wave period, as well as the skewness (*Sk*) and asymmetry (*As*) parameters that characterize nonlinear wave shape (Eq. (10)). For the four example locations shown, which span the shoaling, outer surf zone, and inner surf zone regions, the $\tilde{\eta}(t)$ timeseries agree well with the observations. This includes reproducing the more complex wave shapes for the plunging case, where there is very rapid rise of the free surface just prior to the arrival of the crest, followed by the much slower fall of the free surface that includes an inflection (small second local maximum) in the water level. While this inflection is slightly more pronounced in the observations, its occurrence and timing are still predicted by the model. The close agreement between the modelled and observed wave skewness and asymmetry is particularly remarkable, given this covers a large number of measurement locations and wave states as they transform from deep water to the shoreline, and given that as third-order bulk statistical parameters of the waves are generally more difficult to predict relative to wave heights. For the spilling case, the model captures the significant increase in *Sk* as the waves steepen in the shoaling zone up the breakpoint ($x_b \approx 6.5$ m), which is followed by a gradual return towards a linear wave shape as the shoreline is approached (Figure 4e). Similarly, the model captures the two inflections (local maxima) in *As* that are observed seaward of the breakpoint and the sharper decrease in *As* (more negative, implying a pitched forward wave face) that occurs shoreward of the breakpoint. While the available data for the plunging case are more limited, which only include measurements shoreward of the breakpoint, the modelled *Sk* and *As* agree similarly well with the observations.

The mean Eulerian current profiles show distinct reversals in flow above and below the mean wave trough elevation consistent with an expected undertow profile, with onshore flow within the crest region and offshore flow below the trough (Figure 5). As noted by TK94, due to restrictions with the LDA measurements they were only able to obtain a small portion of valid data above the trough; however, they measured the entire water column beneath it. The modelled mean current profiles are generally in good agreement with the observations. Above the trough, the model compares reasonably well in the lower portion of the region where measurements could be made; further above this elevation (where no observations are available), the model predicts a local maximum in each current profile with the velocity then decreasing again towards the crest. Below the trough, for the spilling case the model tends to accurately predict the shape and magnitude of the undertow profile; however, at some locations shoreward of the breakpoint (particularly those at *x*=7.0-8.5 m), the model tends to predict a slightly more vertically-sheared profile. For the plunging case, the undertow within the inner surf zone (*x*>8.5 m) is also accurately predicted; however, similar to the spilling case, within the outer surf zone immediately seaward of the breakpoint (i.e., locations *x*=7.0 – 8.5 m), there is marginally more vertical shear in the undertow profile than observed and overall there is slightly too much flow offshore.

### 3.2 Wave breaking over a fringing reef (Yao et al. 2012)

The two Y12 cases have similar offshore wave conditions but use different offshore still water levels (Figure 6). In both cases, strong plunging waves break over the steep forereef slope; however, the difference in still water level greatly influences how rapidly the waves break and dissipate energy within the surf zone. For Case 1, where the still water depth over the reef is $h_r$=0.1 m, wave breaking occurs near the crest (*x*=0 m), and after a sequence of splash-up events, a bore propagates shoreward that allows wave energy to be transmitted across the reef flat. For Case 3, the still water depth over the reef is $h_r$=0.0 m; however, due to wave setup, there is still on average a small depth of water over the reef flat (~0.04 m). Due to the shallow depth, for this case the wave drawdown preceding breaking causes the forereef slope to be nearly dry (minimum ensemble-averaged depth of ~2 cm), causing the plunging jet to impinge on the bottom, leading to a large splash-up of water. Nearly all of the incident wave energy is dissipated near the crest region, with the presence of only a small depth-limited bore propagating across the reef flat.

Y12 recorded video imagery of the surf zone region for Case 1, and while the imagery is not rigorously georeferenced and there can be variability between individual breaking waves, it can at least be qualitatively compared to the free surface profiles predicted by the model. Figure 7 shows an image sequence of an individual breaking wave with the free surface predicted by the model superimposed. Note that even with regular wave breaking there is some expected variability in how individual waves break (i.e., due to some element of chaotic behaviour as described by e.g. Wei et al. (2018)) and the video was not synchronized with the wavemaker or instrumentation. Therefore, in Figure 7 we compare the breaking of an arbitrary wave, which tends to reproduce the main features of the breaking process well, including the initial plunging wave shape, an initial splash-up, and subsequent propagation of the bore across the reef flat.

The model accurately predicts the ensemble-averaged wave crest ($\tilde{\eta}_{max}$) and trough ($\tilde{\eta}_{min}$) elevations for both cases (Figure 8). The wave setup ($\bar{\eta}$) profiles are also reproduced reasonably well, with good agreement with the maximum setup generated over the reef flat, including the much greater setup (factor of ~3) for the shallower depth Case 3. For Case 1, the model predicts less setdown in the shoaling zone near the crest (Figure 8c), and for Case 3, does not predict the jump in wave setup further back on the reef flat at $x \approx 1.7$ m (Figure 8d). For Case 3, this significant jump in setup (~20%) would likely to be an experimental artefact, as there is effectively no wave energy remaining at this location on the reef (and hence no radiation stress gradients) to explain a sudden generation of wave setup. For Case 1, the cause of the discrepancy in setdown is not clear, although we note that Y12 found that setdown was similarly underpredicted using a Boussinesq wave model simulation applied to this same case.

The model predictions of wave shape generally agree well with the observations, as illustrated by the comparison of the ensemble-averaged water level timeseries at four locations, as well as the agreement with the skewness and asymmetry properties over the whole domain (Figure 9). For Case 1, the predicted water level timeseries are nearly in exact agreement with the observations (Figure 9a-d). For Case 3, the predicted water level timeseries agree well at locations offshore of the crest (*x*<0) (Figure 9g-i); however, there is slightly poorer agreement on the reef flat, albeit at this location the wave heights are minimal ($H \approx 0.006$ m), and hence more than an order of magnitude smaller than offshore. For Case 1, the skewness (*Sk*) and asymmetry (*As*) patterns across the reef profile are rather striking, with a series of oscillations that are present in both the observations and model predictions. For Case 3, there is poorer agreement between the model predictions and observations, particularly over the reef flat (*x*>0 m); however, the wave heights over the reef flat are very small (order 1 mm) and thus would exaggerate any small differences between the model and observations.

To understand the causes of the oscillations in wave skewness and asymmetry across the reef flat, we further investigate how individual waves change in shape as they propagate shoreward from the surf zone (Figure 10). Eldeberky and Battjes (1995) observed analogous reversals in the skewness and asymmetry of waves during experiments investigating how waves transform as they propagate across a trapezoidal bar, which they attributed to the release of bound super-harmonic waves that, once free, move with different phase speeds and then alter wave shape as they interact. The wave spectrum at the seaward portion of the reef flat ($x=2$ m) shows that there is significant wave energy in the first two super-harmonics ($f_1$ and $f_2$) (Figure 10a), which as free waves with different frequency $f$ would propagate at different speeds $c(f)$, even in the shallow water depth over the reef flat. This is illustrated in Figure 10b, where the wave propagation pathlines $x(t) = c(f)t$ based on linear wave theory predict that both $f_1$ and $f_2$ move slower than a wave with the incident fundamental frequency ($f_0$). Evidence for the release of super-harmonics is particularly clear for $f_2$ in Figure 10b, where a distinct separate wave crest moves at the phase speed predicted by the second harmonic. Initially this wave contributes to steepening the front face of the wave (hence negative $As$) but once it interacts with the next wave of frequency $f_0$ shoreward on the reef flat it generates a steeper back face (hence positive $As$). This swap in wave shape is also clearly visible in the ensemble-averaged water level profile across the reef flat shown in Figure 10c.

Y12 only conducted mean current profile measurements for Case 1, which were obtained only below the wave trough level (Figure 11). The model generally reproduces the main features of the mean current profiles across the reef reasonably well. An interesting feature of the current profiles is the reversal of the undertow profile below the trough that occurs on either side of the reef crest (i.e., increasing towards the bottom shoreward of the reef crest; decreasing towards bottom seaward of the reef crest), which is reproduced by the model.

## 4   Discussion

### 4.1 Wave transformation and wave setup

The cross-shore variations of wave heights and wave shape were accurately reproduced by the SPH simulations over the range of different wave conditions and bathymetry profiles that were considered in this study. This included a broad range of different wave breaking types, spanning the extremes of spilling waves breaking on a plane beach in TK94 (Figure 3) to the strongly plunging waves breaking on a nearly dry, steep reef slope in Y12 (Figure 8). The excellent agreement with the cross-shore evolution of individual wave shape was particularly notable, given how the test cases included complicated and variable changes in the skewness and asymmetry properties of the waves as they propagated across the bathymetry profiles (Figure 4, Figure 9). For example, this was illustrated by the accurate reproduction of the complex skewness and asymmetry oscillations across the reef profile in Y12 (Figure 9), which was achieved by correctly resolving the release of super-harmonics within the surf zone that then disperse with different celerities across the reef platform (Figure 10). Perhaps most significant overall, the SPH simulations were able to reproduce the complex wave transformation processes using a fixed set of numerical parameters using recommended / default values, which required no tuning among the different cases despite the very different wave breaking characteristics.

Nevertheless, it is also important to acknowledge that other phase-resolving modelling approaches based on fixed meshes have also been successful in reproducing wave transformation over a similar range of wave breaking extremes (including for the same case studies considered here); however, these models often require tuning of empirical parameters that can reduce their general predictive skill. As (quasi) depth-averaged versions of these models (e.g., based on Boussinesq and non-hydrostatic approaches) cannot resolve the overturning of the free surface and details of wave breaking, they require some parameterization to describe the breaking process. For example, within Boussinesq models this has commonly been based on the inclusion of empirical roller models (e.g., Schäffer et al., 1993, Madsen et al., 1997) or eddy viscosity formulations (e.g., Cienfuegos et al., 2010, Kennedy et al., 2000); whereas, non-hydrostatic models have tended to use momentum-conserving shock-capturing schemes that treat wave breaking similar to a hydraulic jump (e.g., Zijlema et al., 2011, Ma et al., 2012). For depth-averaged (i.e., single layer) and coarse multi-layer (less than about ten vertical layers) non-hydrostatic models, the acceleration of the crest region that triggers breaking and subsequent development of saw-tooth bores cannot be accurately captured; so these models often use a "hydrostatic front approximation" (HFA) when a critical wave steepness locally occurs that quickly transitions the surface into a bore-like shape that dissipates energy (Smit et al., 2013). Despite the relative simplicity of how the wave breaking physics within (quasi) depth-averaged models are described, they have nonetheless been able to reproduce many of the characteristics of wave transformation, including for model applications to the same test cases, i.e., for both TK94 (Roeber et al., 2010, Bredmose et al., 2004, Lynett, 2006, Tissier et al., 2012, Cienfuegos et al., 2010, Derakhti et al., 2016a, Bradford, 2010) and Y12 (e.g., Yao et al., 2012, Zhang et al., 2018b). However, to achieve this requires adjusting empirical parameters that affect wave energy dissipation within these models (including the onset and cessation of breaking) (e.g.,

Smit et al., 2014), which can vary for different breaking wave conditions and hence does not strictly have a strong physical basis.

Three-dimensional (3D) phase-resolving wave-flow models that resolve the vertical structure, on the other hand, can provide a more complete description of the wave breaking process. For the case of CFD models based on RANS or LES (for example, applications of OpenFOAM), they are able to simulate overturning breaking waves on a fixed mesh, including surf zone turbulence. While computationally expensive, these approaches have been very successful in reproducing a wide range of nearshore hydrodynamic processes observed in experimental studies. For example, a number of studies have similarly compared CFD models to the experimental observations of TK94, and found that these approaches can accurately reproduce surf zone wave transformation and wave setup profiles with comparable skill to the present study (e.g., Jacobsen et al., 2012, Brown et al., 2016, Larsen and Fuhrman, 2018). Notably, these studies have suggested that predictions of wave transformation and setup profiles are not particularly sensitive to having very accurate reproduction of the surf zone turbulence. For example, Brown et al. (2016) assessed 6 turbulence closure schemes and found that all could reproduce the surface elevations observed in TK94 (see Figure 2 in that paper), despite the significant differences in the mean and turbulent flow structures predicted by these approaches (see further discussion below).

Multi-layered nonhydrostatic models such as SWASH (Zijlema et al., 2011) and NHWAVE (Ma et al., 2012) were developed to be more computationally efficient alternatives to CFD models, albeit they describe the free-surface with a single value and thus do not attempt to reproduce the overturning that occurs during breaking. Nevertheless, despite this rather crude representation of the breaking physics, with sufficient vertical resolution, these models have proven to be remarkably robust at reproducing wave transformation and setup profiles over a wide range of wave conditions (including plunging), without the need for use of the HFA to trigger the onset of breaking (and hence avoiding the need to specify an empirical parameter to prescribe the maximum wave steepness). For example, Smit et al. (2013) investigated how multi-layered applications of SWASH performed at simulating the TK94 experiments with the HFA disabled. They found that with 20 layers the initiation of breaking could be accurately captured even for plunging waves, resulting in robust predictions of both the wave transformation and setup distributions, independent of a detailed resolution of the breaking process.

Overall, it appears that for nearshore applications where the primary aim is to accurately predict the water level variations through the surf zone (not necessarily details of the mean and turbulent flow structure), the use of a nonlinear phase-resolving model is of utmost importance. However, models with finer vertical resolution (i.e., 3D or 2DV) are required to capture wave breaking without calibration, as depth-averaged or coarse vertical resolution Boussinesq-type and non-hydrostatic models rely on some sort of breaker model (that either accounts for the dissipation, such as the eddy viscosity and roller approach, or triggers the initiation of wave breaking, such as the HFA approach). Therefore, the results suggest that both mesh-based (i.e., CFD, non-hydrostatic and Boussinesq-type) and mesh-free (i.e., SPH) models can all be equally capable of delivering accurate predictions of wave transformation and setup despite differences in how they reproduce the detailed characteristics of breaking waves.

### 4.2 Wave-driven transport and surf zone turbulence

The SPH simulations provided reasonably accurate representations of the cross-shore variations in the mean flow structure observed within both the TK94 (plane beach) and Y12 (reef profile) experiments. For the TK94 experiments, the magnitude of the undertow strength was generally well predicted, albeit with some discrepancies observed in the vertical flow profiles at some cross-shore locations. For both the spilling and plunging cases (Figure 5), the model results displayed some remarkable agreement with the observations, particularly within the inner surf zone ($x > 8.5$ m); there were only some small discrepancies observed in the outer surf zone immediately near or shoreward of the breakpoint (7.0 m $< x <$ 8.5 m) where the modelled profiles were slightly more sheared. For the available flow measurements above the mean wave trough elevation, the simulations accurately reproduced the onshore mean transport profile within the crest region (Figure 5) and given the good agreement with offshore volume flux below the trough, an equally good agreement with the onshore volume flux within the crest region that could not be measured can also be inferred.

Given that the accurate prediction of cross-shore transport in the nearshore is critical to a wide range of applications (e.g., understanding how waves drive beach erosion and recovery cycles), several studies have investigated the performance of numerical models to reproduce experimental observations of wave-driven mean flows. These studies include those using 3D / 2DV phase-resolving wave-flow models solved on fixed grids using the same TK94 experiments to benchmark model performance, which can therefore provide valuable context to the present SPH results. Rijnsdorp et al. (2017) and Derakhti et al. (2016a) applied the multi-layered non-hydrostatic model approach to evaluate its ability to reproduce the mean current profiles observed in TK94 (see Figure 6 in the first, and Figure 3 and 4 in the latter study). While both studies found that the non-hydrostatic approach could reproduce the main features of the current profiles reasonably well, the results deviated significantly more than in the present study, with both an overprediction of the onshore transport in the crest region (by a factor of 2 for the plunging case) and with a similar overprediction of the offshore undertow transport (i.e., defined by more exaggerated belly-shaped profiles). Several studies have also applied CFD models (i.e., OpenFOAM) to the TK94 experiments, and have often also observed a significant overprediction of the undertow transport compared to the present SPH results, particularly when using conventional RANS-based turbulence closure models (e.g., Jacobsen et al., 2012, Brown et al., 2016, Larsen and Fuhrman, 2018, Devolder et al., 2018). Brown et al. (2016) were able to improve predictions of the undertow profiles by adopting more sophisticated turbulence closure schemes (for example, see Figures 5 and 8 in Brown et al. (2016) for comparisons with the spilling and plunging cases, respectively); however, in all cases the model overpredicted the undertow velocities more than in the present study, especially within the inner surf zone region where agreement with the SPH results were excellent. Larsen and Fuhrman (2018) observed similar overpredictions of the undertow in their CFD simulations, which they believed was due to surface rollers travelling too fast due to local underpredictions of eddy viscosities (hence flow resistance) simulated within the upper-most part of the water column, leading to increased onshore flows above the wave trough and increased return flows below the wave trough. While the undertow profiles were generally accurately predicted using the present SPH approach, the small discrepancies observed could likewise be due to some deviations in the vertical turbulent flow structure predicted within the inner surf zone.

      The ability of the present SPH simulations to more accurately predict the mean current profiles in TK94 relative to a number of CFD model applications is somewhat surprising, given that no sub-particle scale turbulence closure model was used; albeit, the simulations themselves were conducted at relatively fine resolution (order 1 mm) where at least many of the larger-scale wave-generated turbulence would be expected to be resolved. However, before discussing the role of turbulence on the current profiles, it should be first acknowledged that some of the improved performance of the SPH results over many prior CFD studies could be due to a more robust representation of the onshore mass transport within the crest region of the breaking waves, which was overpredicted in all of the aforementioned studies (in contrast to present SPH simulations where there was generally excellent agreement). Given that any overprediction of onshore mass transport would lead to an overprediction of the undertow transport below the trough, independent of how vertically distributes the momentum, at least some of the strong performance of the SPH simulations could simply be better resolution of the onshore mass flux in the crest region during breaking. This might be expected when comparing to the case of a non-hydrostatic model where a single-valued representation of the free surface cannot truly resolve overturning wave breaking, potentially leading to some discrepancy in onshore mass flux. However, CFD models based on VOF are arguably able to resolve the overturning free surface, albeit at a resolution prescribed by

the numerical mesh (ranging from 0.005 – 0.01 m in the aforementioned studies). Further work is required to understand the capabilities and limitations of different modelling approaches (both fixed-mesh and mesh-free) to accurately predict mass transport within the crest region of breaking waves; this would also be best investigated using additional experimental datasets with more observations than TK94, which only reported observations within the lower portion of the wave crest.

Once a model simulates a given onshore mass flux within the wave crest, the properties of the surf zone turbulence then influence the vertical distribution of the undertow profile. Given that surf zone turbulence spans a broad range of scales, the SPH simulations would include motions that are directly resolved (i.e., at scales greater than the averaging kernel) but would neglect smaller (sub-particle) scale that would only be parameterised using a simple viscosity formulation in the present study. As defined in Eq. (8), for regular waves it is conventional to decompose an instantaneous velocity timeseries $u(t)$ via ensemble averaging into a mean ($\bar{u}$), wave ($\tilde{u}(t)$), and turbulent ($u'(t)$) motions; in this sense, turbulence is defined as any unsteady motions that are non-periodic (Nadaoka et al., 1989). While a detailed investigation of the surf zone turbulence predicted by the SPH simulations is beyond the scope of the present study, and would also be best complemented by detailed assessments of sub-particle scale (SPS) turbulence closure models that have been implemented in SPH, it is nonetheless interesting to assess what the present high-resolution model simulations predict using the turbulent motions that are directly resolved within the resolution of the simulations. Figure 12 compares the mean turbulent kinetic energy (TKE) $\bar{k}_t$ reported in TK94 for the spilling and plunging cases with the model predictions. Note that for comparison we adopt the identical analysis approach presented in TK94, where given that transverse flows were not measured (and also not simulated in the 2DV simulations), the wave phase-varying TKE was calculated as $k_t = (1.33/2)\left(\overline{u'^2} + \overline{w'^2}\right)$ and then time-averaged to obtain $\bar{k}_t$. For both the spilling and plunging cases, the model agrees reasonably well with the observations within the inner surf zone region (i.e., $x$>9 m and 10 m for the spilling and plunging cases, respectively); however, within the outer surf zone towards the break point, the model overpredicts the mean TKE. A number of CFD and non-hydrostatic model applications of TK94 have also compared TKE predictions (e.g., Jacobsen, 2011, Larsen and Fuhrman, 2018, Rijnsdorp et al., 2017, Brown et al., 2016), and have similarly found TKE to be significantly overpredicted when using conventional turbulence closure schemes within RANS-based models. This has motivated the application of more sophisticated closure models (e.g., Devolder et al., 2017, Brown et al., 2016) that consider, for example, the stabilizing effects of buoyancy due to entrained air, which has helped to improve model agreement; nevertheless, it is still widely recognized that there is considerable scope to further improve parameterization of sub-grid scale turbulence within surf zone applications within these models.

For the present SPH simulations, it is interesting that the computed TKE was overpredicted even with no turbulence closure (i.e., SPS) model included. Within SPH, turbulent motions can, in general, be described as those that can be directly resolved at the particle resolution (i.e., $>h_{SPH}$) and those sub-particle scale motions that may be parameterized within a turbulence model (if included within an application) (e.g., Dalrymple and Rogers, 2006). Therefore, analogous to how turbulence is described in fixed-grid CFD models (i.e., either RANS or LES), where the TKE should be a superposition of the turbulent motions resolved on the grid and the sub-grid scale turbulence parameterized by the closure model, the inclusion of an SPS closure model in the present study would potentially result in an even greater over prediction of the total TKE. While the source(s) of the overprediction of TKE in the present SPH simulation are unclear, and may be due to both the two-dimensional nature of the simulations and the simplified treatment of viscous stresses in the present study, at least another potential candidate could include the absence of multi-phase behaviour within the simulations, where in reality the buoyancy of entrained air during breaking would be expected to have some stabilizing effect on any turbulence generated, which has been identified in various CFD studies of surf zone turbulence (e.g., Devolder et al., 2018).

Given that the model appears to resolve a significant amount of turbulence generated within the surf zone, we can also assess how the spectral properties of the turbulent velocity fluctuations compare with known trends reported in the literature. Figure 13 shows power spectra for both horizontal ($S_{u'u'}$) and vertical ($S_{w'w'}$) turbulent velocity components at two locations (outer and inner surf zone) and at two elevations (mean trough elevation and near the bottom) for the spilling case. Within these figures, there are regions where the spectra display both -3 and -5/3 slopes (in log-log space) that are commonly observed within surf zone turbulence measurements. Within the outer surf zone where incipient breaking occurs ($x$=7 m), at the mean trough elevation there is a well-defined -3 slope for $f$ >1 Hz (Figure 13a). Near the bottom (Figure 13b), there is also a -3 slope between $1 < f < 5$ Hz; however, at higher frequencies ($f > 5$ Hz) a -5/3 slope is observed. Within the inner surf zone ($x$=10 m), at the mean trough elevation there is also a broad region ($f > 3$ Hz) with a -3 slope (Figure 13c). Near the bottom, the horizontal velocity spectrum ($S_{u'u'}$) displays both -3 and -5/3 sloping regions (Figure 13d); however, the vertical velocity spectrum ($S_{w'w'}$) displays a -5/3 slope over a broad range ($f > 0.5$ Hz).

A number of experimental studies have investigated the spectral properties of surf zone turbulence, which have frequently found that the spectral shapes are initially dominated by a -3 slope close to the breaking point (both in cross-shore location and towards the free surface), which transitions to increasingly -5/3 slopes as the wave breaking evolves shoreward (e.g., Stansby and Feng, 2005, Lakehal and Liovic, 2011, Hattori and Aono, 1985, Lemmin et al., 1974, Battjes and Sakai, 1981). Ting and Kirby (1996) report the turbulent velocity spectra for the TK94 spilling case, albeit at only one cross-shore location within the inner surf zone ($x$=10.4 m) and within the lower portion of the water column (both near the bottom and at intermediate depth between the trough and bottom), where a dominant -5/3 spectral slope was observed for $f > 1$ Hz. This location is most closely related to the model results presented in Figure 13d, where the vertical turbulent velocity spectrum ($S_{w'w'}$) also displays a broad -5/3 slope for $f > 1$ Hz; however, for the horizontal component ($S_{u'u'}$) the model predicts a narrower -5/3 region confined to higher frequencies ($f > 4$ Hz), with a steeper -3 slope observed at intermediate frequencies ($1 < f > 4$ Hz). While the cause of this discrepancy is not entirely clear, this would likely be due to the 2D representation of turbulent motions in the present simulations that would likely not adequately describe the turbulent energy cascade that would be present within experiments. For example, based on theory of 2D homogeneous turbulence, an inertial sub-range with a -5/3 spectral slope is predicted to occur at small spatial-scales, thus similar to the inertial sub-range for 3D turbulence (Lesieur, 2008). However, for 2D the direction of energy flow within this inertial sub-range can be from small to large scales, thus opposite to the direction with 3D turbulence. A detailed investigation of surf zone turbulence, including the role of 2D versus 3D turbulent motions, is well beyond the scope of the present study but should be conducted in the future. However, given that the present 2D model results are qualitatively consistent with trends in experimental observations of surf zone turbulence, this suggest that at least some of the important turbulent motions can be resolved by 2D simulations.

**4.3 Implications for improved understanding and predictions of surf zone processes**

Overall, the results indicate that the SPH approach can be used as a robust and powerful tool to simulate detailed surf zone hydrodynamic processes, with comparable model skill to state-of-the-art mesh-based CFD models. This conclusion is analogous to recent work by González-Cao et al. (2019) who evaluated the ability of both a mesh-free (DualSPHysics) and mesh-based (OpenFOAM) CFD models to simulate wave impacts on coastal structures, and found that both models reproduced the experimental observations with similar accuracy.

One potential strength of SPH models is their inherent suitability for simulating wave breaking (including overturning plunging waves) where the free surface displacements are extremely variable, and hence do not naturally conform to a fixed mesh. The results suggest that a mesh-free SPH approach can provide robust predictions of surf zone wave breaking and help to overcome some of the inherent challenges with quantifying hydrodynamic processes near the free surface (in particular within the critical crest-to-trough region) that are extremely difficult to measure experimentally. On this basis, further analysis of the SPH model results should help to provide new insight into surf zone processes (including momentum and energy balances) beyond conventional phase-resolving wave models. Therefore, while a comprehensive investigation of these surf zone processes is beyond the scope of the present study, here we highlight some additional aspects of the results that are particularly relevant to understanding wave transformation and momentum balances in the nearshore.

As an example, we can further examine the results for the spilling case of TK94, where a region of elevated positive (clockwise) vorticity emerges within the crest region as each wave breaks (Figure 14). This concentrated region of vorticity coincides with formation of a roller (Svendsen, 1984), where potential energy within the wave is initially converted to organized kinetic energy, prior to dissipation occurring. As the roller propagates shoreward within the wave crest, it leaves behind a series of surf zone eddies (rotating both clockwise and counter-clockwise). Pairs of eddies with counter-rotating vorticity appear to cause water within the crest to be advected downwards towards the seabed (i.e., sequence from Figure 14c to d), analogous to the descending eddy pairs that have been observed in detailed experimental and numerical studies of breaking waves (e.g., Nadaoka et al., 1989, Farahani and Dalrymple, 2014, Zhou et al., 2014). These eddies also persist to some degree over a full wave cycle, with a following wave interacting with residual eddies that were previously generated during breaking (i.e., Figure 14a).

The spatial variability in the water levels and flows (mean and turbulent) generated during breaking ultimately determine the mean (wave-averaged) momentum balances that govern wave setup distributions and mean currents through a surf zone. Given that the SPH approach should resolve additional physics relative to conventional phase-averaged and phase-resolving wave models, we can further interrogate the results to examine how various processes that are generally neglected (or parameterized) in conventional models influence wave transformation, and in turn how they influence predictions of cross-shore momentum balances in the nearshore. In the absence of bed shear stresses, the depth- and time-averaged momentum equation is (e.g., Mei et al., 2005)

$$U \frac{\partial U}{\partial x} = -g \frac{\partial \bar{\eta}}{\partial x} - \frac{1}{\rho (\bar{\eta} + h)} \frac{\partial S_{xx}}{\partial x} \qquad (11)$$

where $U$ is the mean (depth and time-averaged) current, $\bar{\eta}$ is the wave setup and $S_{xx}$ is the radiation stress representing the excess momentum flux associated with waves (including turbulence contributions). The response of mean water levels and flow in the nearshore thus depend on how radiation stresses evolve through the surf zone per Equation (11). The radiation stress ($S_{xx}$) can be defined in general form as (e.g., Mei et al., 2005)

$$S_{xx} = \int_{-h}^{\eta} \overline{\left(\rho \tilde{u}^2 - \rho \tilde{w}^2 + \rho \overline{u'^2} - \rho \overline{w'^2}\right)} dz + \frac{1}{2} \rho g \overline{\eta^2} \quad (12)$$

$$\underbrace{\phantom{\rho \tilde{u}^2 - \rho \tilde{w}^2}}_{\text{wave part}} \underbrace{\phantom{\rho \overline{u'^2} - \rho \overline{w'^2}}}_{\text{turbulent part}}$$

which reveals that it is comprised of terms related to both wave and turbulent velocities, as well as the water level variance. Therefore, a primary challenge in surf zone modelling is how to accurately account for various contributions to radiation stresses in the presence of nonlinear, breaking waves. Phase-averaged wave models are based on linear wave theory (LWT), where the radiation stress is given as

$$S_{xx,LWT} = \left(2n - \frac{1}{2}\right)E \quad (13)$$

where $n = c_g / c$ represents the ratio of the group ($c_g$) and phase ($c$) velocities, $E = \rho g \overline{\eta^2}$ is the energy density with PE and KE contributions being in balance for linear waves. These phase-averaged models are often supplemented by empirical corrections to parameterize nonlinear wave behaviour as well as the rate at which PE is converted to KE and then dissipated within breaking waves (i.e., through use of empirical roller models). Conventional phase-resolving models (e.g., Boussinesq and non-hydrostatic) resolve the nonlinear wave characteristics that influence radiation stresses; however, they cannot fully resolve the wave breaking processes that can also affect radiation stress estimates (particularly through robust predictions of surf zone turbulent motions).

To explore how different components of the radiation stress vary across the surf zone, we decompose the individual terms in Eq. (12) that contribute to the cross-shore variability in $S_{xx}$ during breaking, again using the TK94 spilling case as an illustrative example (Figure 15). The results reveal the significant role that the turbulent contribution plays (following breaking) in shifting the radiation stress gradients shoreward (Figure 15a), and hence the wave forces responsible for generation of wave setup via Eq. (12). While the dominant wave contribution to the radiation stress ($\tilde{u}^2$) decays monotonically during breaking, this triggers a rise in turbulent contribution ($\overline{u'^2}$) that shifts the total (combined) $S_{xx}$ shoreward (Figure 15b). In phase-averaged wave models, as well as depth-averaged phase-resolving models, this delay in wave forces can only be parameterized, which for example, historically motivated the development of roller models to improve predictions of wave setup and wave-driven flows within the surf zone (Svendsen, 1984). The results reveal how the present SPH simulations can incorporate the detailed hydrodynamic processes responsible for generating wave forces within the surf zone, which can explain why the model was able to accurately predict both wave setup and mean current distributions for all test cases considered.

We can lastly compare how the radiation stresses ($S_{xx}$) deviate from linear wave theory (LWT), given that it still forms the basis of coastal-scale wave predictions using phase-averaged models. In particular, following Svendsen and Putrevu (1993) we define the wave parameter $P$ by normalizing the radiation stress as $P = S_{xx} / \rho g H^2$ (Figure 15c), where the wave height $H$ is equivalent to $H = \sqrt{8\overline{\eta^2}}$ (Torres-Freyermuth et al., 2007). Therefore, given that the waves in this example are shallow, LWT holds when $P = 3/16 \approx 0.19$ (as indicated by the horizontal dashed line in Figure 15c). The results indicate that in the shoaling region $P \approx 0.14$, thus significantly lower than LWT would predict; however, following breaking $P$ increases significantly above LWT, reaching values up to $P \approx 0.22$ as a consequence of the enhancement of the radiation stress by surf zone turbulent kinetic energy. These results are very similar to the experimental findings of Svendsen and Putrevu (1993), who examined experimental observations of radiation stresses on mild-sloping plane beaches, and found that $P$ was often much less compared to what LWT would predict during shoaling due to the nonlinear shape of waves (i.e., skewness, associated with the waves having narrow, peaked crests and long, shallow troughs). Similarly, they observed that $P$ reached elevated values following breaking (typically $P$=0.2-0.3), which is also consistent with the present results.

## 5 Conclusions

This study has demonstrated how the mesh-free SPH approach can provide accurate and robust predictions of complex surf zone hydrodynamic processes generated by wave breaking, with model performance comparable to applications of state-of-the-art mesh-based CFD models such as OpenFOAM. Over the wide range of wave breaking types considered, the SPH approach was able to reproduce many of the detailed hydrodynamic processes that govern the nonlinear evolution of wave shape in the nearshore, the free surface characteristics of breaking waves (including violent, plunging waves), the processes governing energy conversion between potential and kinetic energy within the surf zone, and the resulting mean wave-driven flow properties (including wave setup and undertow profiles). A particular advantage of approach used here (the weakly-compressible SPH code DualSPHysics), was its ability to run on computationally-efficient GPUs that enabled high-resolution simulations (sub-millimetre particle spacing) of the experimental results to be achieved on a single-GPU.

Given that the performance of the SPH approach was evaluated using common experimental test cases (e.g., TK94) that have been widely-applied to benchmark the performance of other classes of phase-resolving wave-flow models (e.g., Boussinesq, non-hydrostatic, and CFD), the results also provided an opportunity to inter-compare how different defining characteristics of these models may influence model performance. Based on prior model applications to these experimental datasets, it is clear that all of these phase-resolving models (including depth-averaged versions) are fully-capable of accurately resolving the nonlinear evolution of individual waves prior to breaking. However, within the surf zone region, where phase-resolving models cannot directly resolve the overturning free surface, more variable model performance has been reported across the literature. By necessity, depth-averaged versions of these models (e.g., based on Boussinesq and non-hydrostatic approaches) require significant empirical parameterization of the breaking process. Therefore, while these models have often been successful in reproducing surf zone wave transformation, they generally require tuning of empirical parameters (generally on a case-by-case basis), which can undermine their broader predictive utility and may also come at the expense of other hydrodynamic predictions (e.g., degrading that accuracy of wave setup distributions).

Mesh-based 3D/2DV models with vertical resolution (e.g., multi-layer non-hydrostatic and CFD models) have been shown in recent years to provide more robust predictions of surf zone wave transformation relative to depth-averaged models, as they are much less dependent on empirical parameterization of the breaking process. Of these models, CFD models based on full solution of the Reynolds Averaged Navier Stokes equations are most analogous to the present SPH simulations as they both can directly resolve overturning breaking waves. In the context of recent applications of high-resolution mesh-based CFD models (i.e., OpenFOAM) that have been applied to the same experimental test cases (e.g., Larsen and Fuhrman, 2018, Brown et al., 2016, Jacobsen et al., 2012), the present results indicate that the SPH approach can reproduce these surf zone processes with comparable skill. In fact, the results of this work suggest that the SPH approach can help to improve predictions within the crest region of breaking waves, as evident by robust predictions of cross-shore mass fluxes and undertow profiles that have been notoriously difficult to predict in mesh-based CFD models. Within the present study, it was also significant to find that the SPH approach was capable of accurately simulating the range of wave breaking conditions across the test cases using a fixed set of model parameters that were consistent with recommended values (i.e. approximately default values within the model). While the range of wave conditions and bathymetry profiles considered are by no means complete in this single study, these findings suggest that the approach can be applied with some confidence for scenarios where experimental data is not available for validation; for example, applied as a valuable tool for designing detailed physical modelling studies or when detailed experimental measurements are not possible to obtain within a study. More broadly, in the context of nearshore wave modelling, arguably the greatest advantage of mesh-free SPH models is also how readily they can deal with complex geometries (bathymetry and topography) that may not readily confirm to a fixed mesh (grid). Therefore, while both SPH and mesh-based CFD models may be used interchangeably in applications with simple nearshore bathymetries (i.e., as in the test cases considered here), the SPH approach may confer some greater practical advantages when simulating nearshore processes with coastal engineering structures or natural bathymetries that form complex geometries.

While the present study has demonstrated the great promise of the SPH modelling approach to improving understanding and prediction of surf zone hydrodynamics, it is important to acknowledge that the present focus has been on investigating the performance under simple forcing (i.e., regular waves) and simple bathymetry profiles. This approach was deliberately chosen as an initial starting point to help isolate the performance characteristics of SPH models in the simulation of surf zone hydrodynamics using a simple set of wave breaking conditions. This present work should provide a foundation for further SPH modelling studies of surf zone hydrodynamics under more realistic conditions, including irregular wave conditions and more complicated nearshore bathymetry profiles (e.g., barred beaches, various reef geometries, etc.) where a wealth of experimental data also exists to investigate model performance. While such studies are achievable today, the primary constraint (particularly for irregular wave conditions) is the requirement for much longer simulations (typically more than an order of magnitude greater), and hence computational demand, which is required to properly resolve the statistical properties of irregular waves. For the SPH code used in the present study (DualSPHysics), this greater computational demand could be partially offset by recent developments in coupling the SPH model with efficient phase-resolving wave models (e.g., Altomare et al., 2015b, Altomare et al., 2018, Verbrugghe et al., 2018), thereby concentrating the focus of the SPH simulations on the immediate surf zone region; as well as plans for multi-GPU functionality in future releases of the code.

Finally, with SPH applications to coastal problems still in their early stages (certainly in comparison to decades of work using mesh-based models), the many areas of active research and develop of the SPH approach will help to further advance surf zone applications into the future. These numerous developments in the SPH approach include, for example, improved boundary conditions (solid boundaries and at the free surface), inclusion of multi-phase behaviour, enhanced numerical optimization (including adaptive refinement of particle resolution), and greater accuracy using approaches such as incompressible SPH (ISPH) (e.g., refer to recent reviews by Violeau and Rogers (2016) and Gotoh and Khayyer (2018)). One area that deserves attention in future SPH studies of nearshore wave dynamics is the role of surf zone turbulence, with advanced turbulence models (sub-particle scale) still being an active area of research. The present study highlights some interesting attributes of the turbulence fields generated by the breaking waves, but only the 2D turbulent motions that are directly resolved at the particle scale (albeit at relatively high resolution in the present study). Future work is required to understand how the incorporation of advanced sub-turbulent closure schemes and a full description of the 3D dynamics may further improve SPH predictions of surf zone turbulence and its influence on a range of nearshore hydrodynamic processes.

## Appendices

### Appendix A – Model parameter sensitivity

A number of SPH studies have investigated how various model parameters influence predictions of the propagation and breaking of waves (e.g., Roselli et al., 2018, González-Cao et al., 2019, De Padova et al., 2014). We refer the reader to these comprehensive sensitivity analysis studies for general background on how various model parameters within the WCSPH approach can influence wave propagation and breaking characteristics. These studies have often focused on the role of three specific parameters: 1) model resolution, via the initial inter-particle distance ($dp$); 2) artificial viscosity ($\alpha$); and 3) the smoothing coefficient (*coefh*). To explore the influence of these model parameters on the present results, we initially conducted a sensitivity analysis using the TK94 spilling case results by varying these parameters over a range of values ($dp$=2-16 mm, $\alpha$=0.002-0.02, and *coefh*=1.0-1.8; Tables A1-A3). From these simulations, we evaluated how the modelled wave height ($H$), setup ($\bar{\eta}$) and mean current ($\bar{u}$) patterns compared with the experimental observations by computing the root-mean-squared error ($RMSE_{expt}$)

$$RMSE_{expt} = \sqrt{\frac{\sum_{i=1}^{N}\left(X_{model} - X_{expt}\right)^2}{N}}, \qquad (A.1)$$

where $X$ denotes the variable to be compared, $N$ is the total number of data points and the subscripts 'model' and 'expt' denote the predicted and observed results, respectively. For presentation, we normalize the $RMSE_{expt}$ by the range in the observed value to compute the normalized root-mean-squared error ($NRMSE_{expt}$)

$$NRMSE_{\text{expt}} = \frac{RMSE_{\text{expt}}}{X_{\text{expt}}^{\max} - X_{\text{expt}}^{\min}}, \tag{A.2}$$

where the superscripts 'max' and 'min' denote the maximum and minimum values, respectively. To specifically investigate the convergence behaviour of the model results to *dp*, we also computed a second measure of error by using the finest simulation results (*dp*=2 mm) as the reference data (i.e. by replacing $X_{expt}$ with $X_{fine}$) to compute $NRMSE_{\text{fine}}$ using Eqs. (A.1) and (A.2), where the subscript 'fine' denotes the finest resolution results. In this way, the scaling behaviour of model accuracy with *dp* becomes decoupled from any potential errors in the experimental measurements.

The raw results of this sensitivity analysis are included in Supplementary Material (Figures S1-S6) for reference, with the bulk model performance statistics summarised here in Tables A1-A3. As the particle distance *dp* decreases, the results converge to the experimental observations (Figure A1, Table A1) with negligible reductions in $NRMSE_{\text{expt}}$ for *dp*≤3 mm. Based on this analysis, the default (finest resolution) simulations can reproduce the experimental measurements with <5% error for wave height and approximately 10% error for both the setup and mean currents (Table A1). To assess the convergence behaviour of the numerics alone, on Figure A1 we also include the response of the $NRMSE_{\text{fine}}$ in which the error is referenced to the finest simulation with (*dp*=2 mm). These results indicate that $NRMSE_{\text{fine}}$ decreases as $NRMSE_{fine} \sim dp^b$, where a linear regression of the results in log-log space gives *b*=1.41, *b*=1.42, and *b*=1.0 for wave height, setup and mean current, respectively (not shown). Note that a number of SPH studies have also investigated how particle resolution influences model performance in wave applications, often placing the results in the context of the number of particles resolved within a given incident (offshore) wave height $H_0$, usually with recommendations to keep the ratio $H_0/dp$ greater than order 10 (e.g., Roselli et al., 2018). The present results indicate that performance gains were rather small for all quantities (wave heights, setup and currents) when $H_0/dp > 25$ (Table A1).

For all wave applications of SPH using an artificial viscosity $\alpha$ that we are aware of, these studies have consistently found an optimal value of order 0.01; hence, $\alpha = 0.01$ is defined by default in DualSPHysics (see section 2.4). A sensitivity analysis was conducted where $\alpha$ was varied by an order of magnitude to investigate its influence on wave heights, setup and mean currents (Table A2). The results indicate that all three quantities are insensitive to $\alpha$ over a large range; only for the lowest value $\alpha$ =0.002 is some deterioration of the results evident. Note that these results may appear at odds with some studies that have shown a stronger influence of $\alpha$ on results; namely, De Padova et al. (2014) who investigated the role of $\alpha$ on wave transformation.

However, De Padova et al. (2014) used much coarser resolution ($H/dp$<5) than in the present study ($H/dp$=63), which can likely explain why their results were much more sensitive to $\alpha$.

Finally, we assessed the role of the smoothing coefficient (*coefh*) on the results (Table A3). Both the setup and mean current results are mostly insensitive to *coefh* variations over the range considered (1.0-1.8). However, we observed some small improvements in wave height predictions for lower values of *coefh*, so *coefh*=1.2 was used in all simulations.

**Table A1.** Sensitivity analysis and model performance characteristics for simulations with varying initial inter-particle distance (*dp*) for the TK94 spilling case. Note that the * denotes the default *dp* value used within the main study.

| | | | | $NRMSE_{expt}$ [-] | | |
|---|---|---|---|---|---|---|
| *dp* [mm] | $H_0/dp$ | Particles | Computation time | Wave height | Setup | Current |
| 2* | 62.5 | 1216373 | 63.5 hr | 0.037 | 0.091 | 0.104 |
| 3 | 41.7 | 544781 | 23.1 hr | 0.036 | 0.104 | 0.105 |
| 5 | 25.0 | 195563 | 8.1 hr | 0.060 | 0.138 | 0.111 |
| 9 | 13.9 | 61556 | 1.9 hr | 0.142 | 0.199 | 0.137 |
| 14 | 8.9 | 25743 | 1.1 hr | 0.177 | 0.217 | 0.163 |

**Table A2.** Sensitivity analysis of the model results with artificial viscosity ($\alpha$) for the TK94 spilling case. Note that the * denotes the default $\alpha$ value used within the main study.

| | $NRMSE_{expt}$ | | |
|---|---|---|---|
| $\alpha$ [-] | Wave height | Setup | Current |
| 0.002 | 0.119 | 0.279 | 0.172 |
| 0.006 | 0.043 | 0.093 | 0.100 |
| 0.010* | 0.037 | 0.091 | 0.104 |
| 0.015 | 0.031 | 0.095 | 0.097 |
| 0.020 | 0.032 | 0.097 | 0.091 |

**Table A3.** Sensitivity analysis of the model results with smoothing coefficient (*coefh*) for the TK94 spilling case. Note that the * denotes the default *coefh* value used within the main study.

| | $NRMSE_{expt}$ | | |
|---|---|---|---|
| *coefh* [-] | Wave height | Setup | Current |
| 1.0 | 0.035 | 0.091 | 0.100 |
| 1.2* | 0.037 | 0.091 | 0.104 |
| 1.4 | 0.100 | 0.096 | 0.134 |
| 1.6 | 0.091 | 0.101 | 0.116 |
| 1.8 | 0.094 | 0.118 | 0.124 |

## Acknowledgements

This work was supported by resources provided by the Pawsey Supercomputing Centre with funding from the Australian Government and the Government of Western Australia. We are grateful for the helpful discussions about this modelling with Marion Tissier, Ad Reniers, Ap van Dongeren, Dano Roelvink, Robert McCall, and Niels Jacobsen and Marcel Zijlema, while RJL was on a sabbatical in Delft, the Netherlands. RJL thanks both Deltares and Delft University of Technology for hosting his sabbatical time in the Netherlands. CA acknowledges funding from the European Union's Horizon 2020 research and innovation programme under the Marie Sklodowska–Curie grant agreement No.: 792370. YY acknowledges support from the National Natural Science Foundation of China [grant number 51679014]. We finally thank the three anonymous reviewers for providing very helpful feedback that improved the manuscript.

# Figures

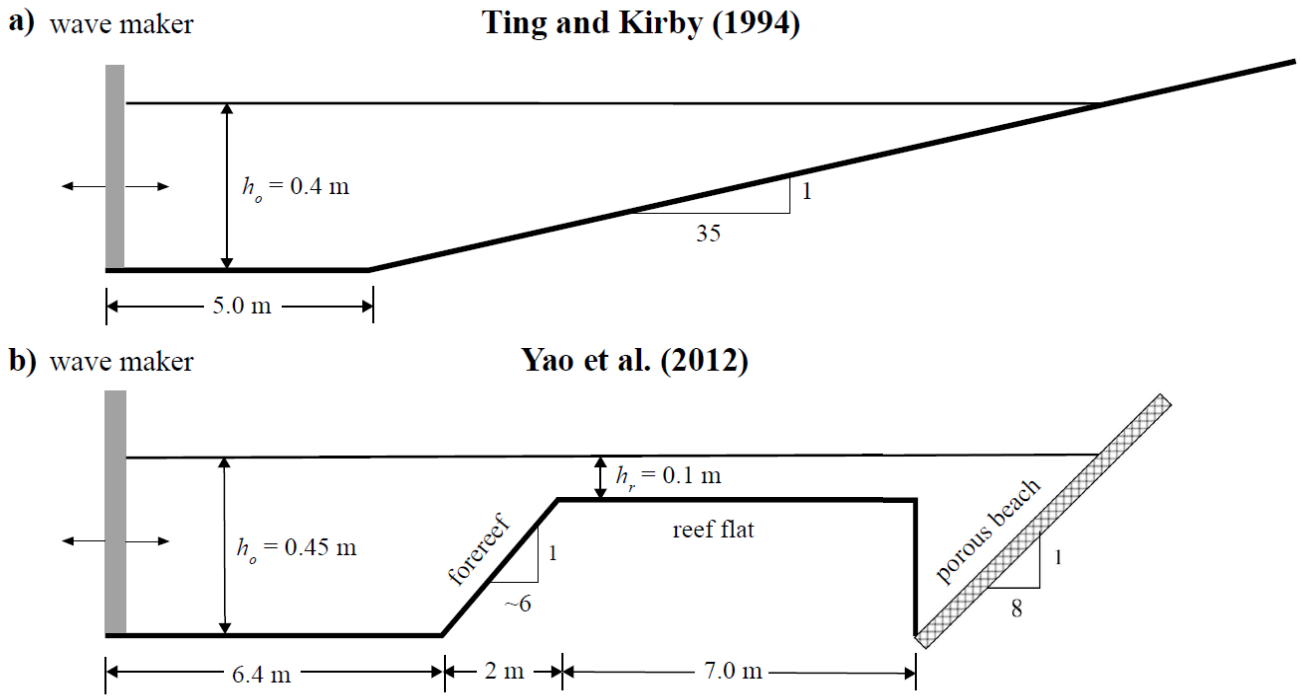

**Figure 1.** Experimental setup as simulated in the model for the a) Ting and Kirby (1994) and b) Yao et al. (2012) test cases. In both cases, the flat offshore region was shortened for computational efficiency (see text for details). The Yao et al. (2012) case is specifically drawn for Case 1 where $h_r$=0.1 m. Note that the vertical scale is exaggerated by 8:1.

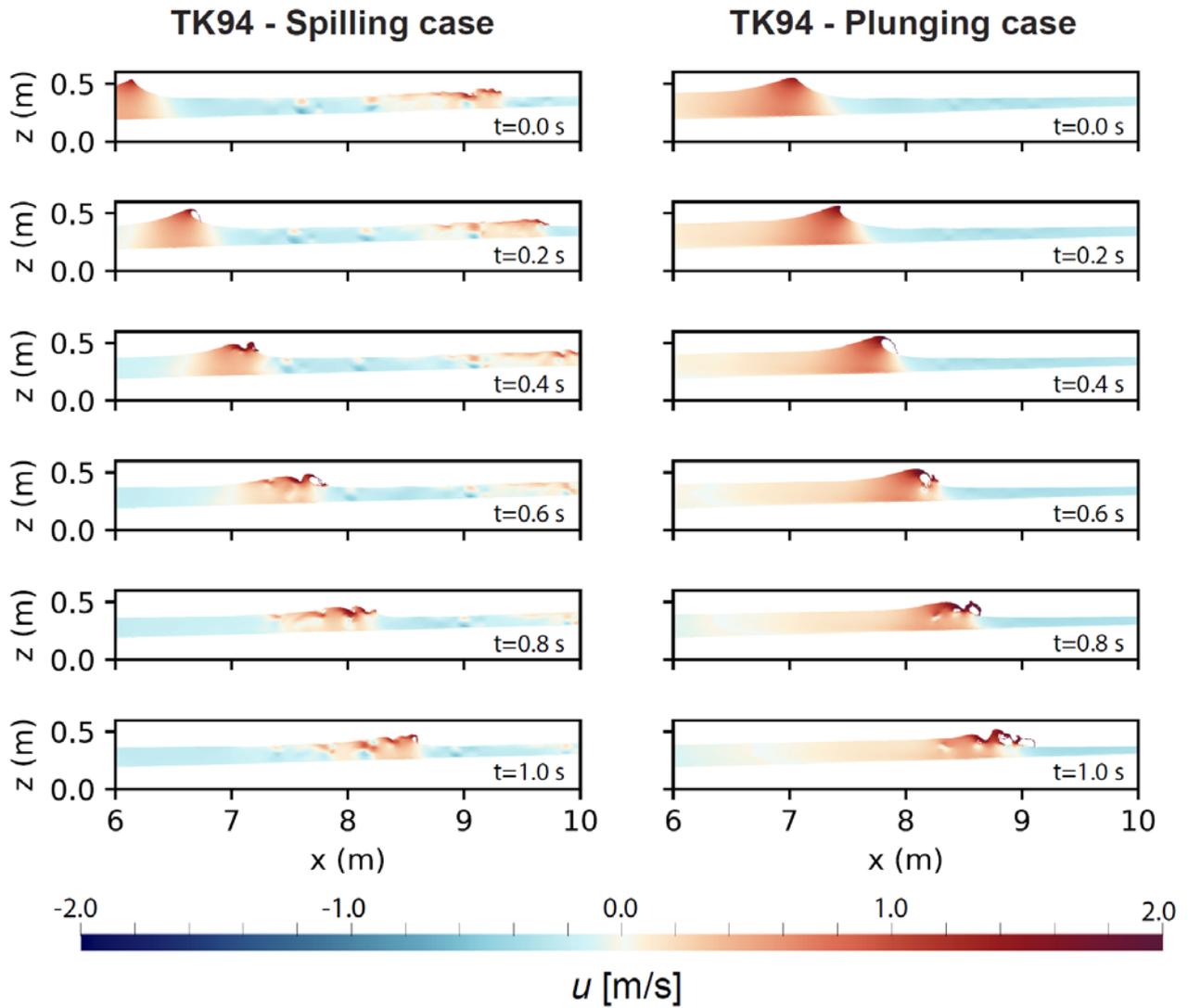

Figure 2. Evolution of wave breaking on a plane beach from TK94 with colours denoting the horizontal ($u$) velocity component. (Left) spilling case. (Right) plunging case. In both cases the velocity output is displayed at 0.2 s interval, where time $t$=0 s is arbitrarily assigned to the first figure in the sequence. Note that due to the shorter period (wave length) in the spilling case, two individual waves are visible within the domain shown.

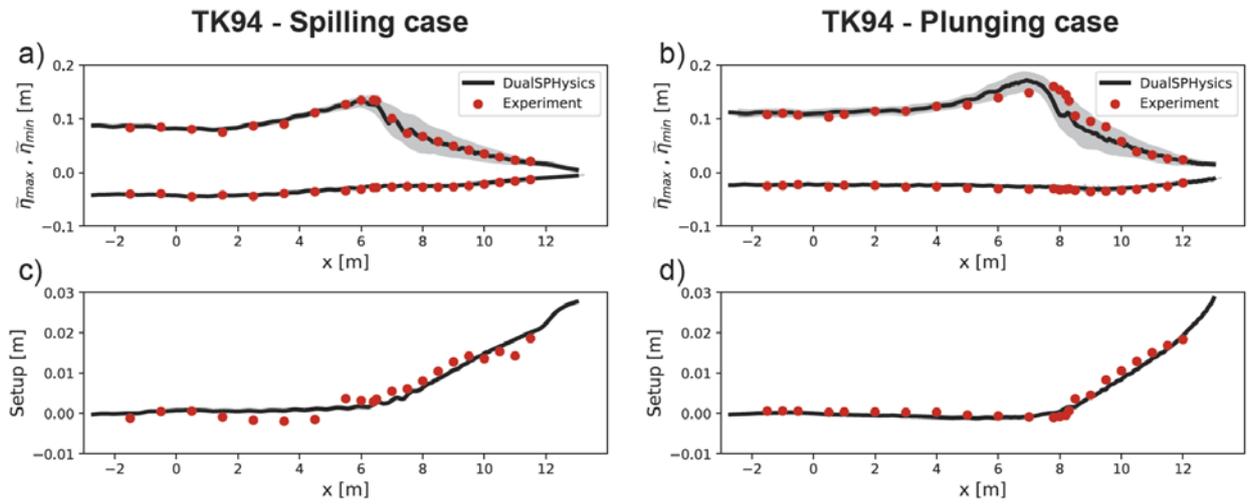

**Figure 3.** Wave height envelope and wave setup evolution for the (Left column) spilling wave case and (Right column) plunging wave case in TK94. (Top row) envelope of the maximum crest elevation ($\tilde{\eta}_{max}$) and minimum trough elevation ($\tilde{\eta}_{min}$) derived from the wave ensemble-averaged water levels. (Bottom row) wave setup. Grey regions denote one standard deviation of the water elevations from the ensemble (wave) average. Note that the standard deviation of the trough elevation is much smaller than the crest elevation, and hence not visible.

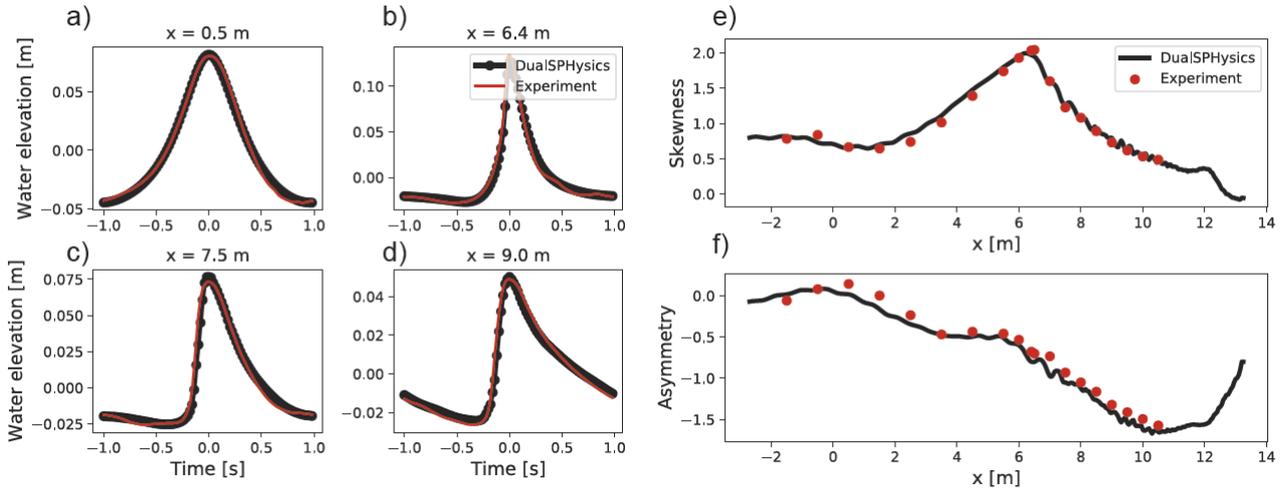

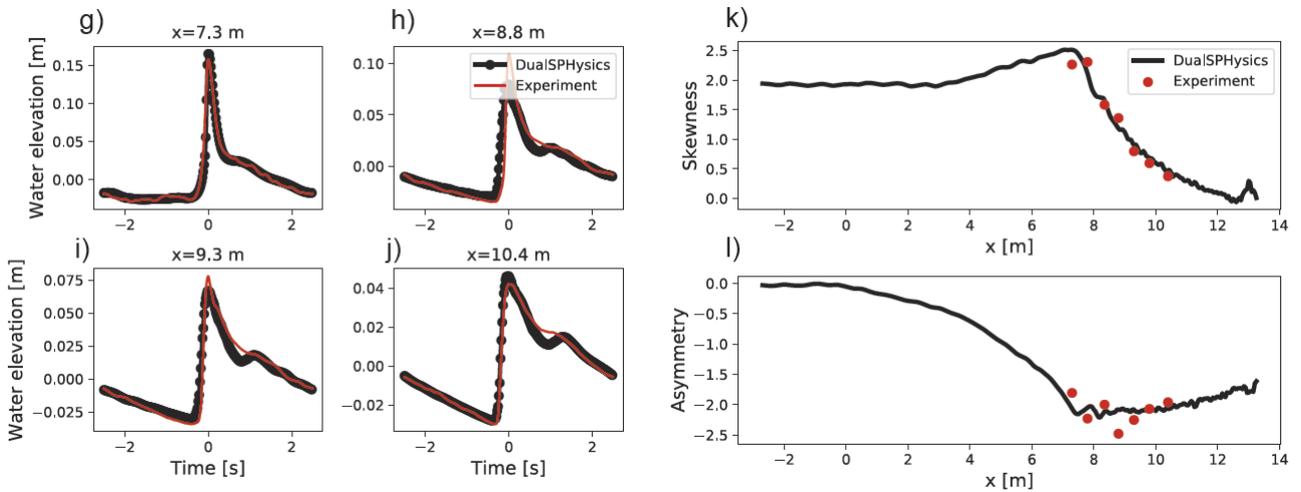

**Figure 4.** Water surface elevation (wave ensemble-averaged) and wave shape (skewness and asymmetry) evolution for the TK94 experiments. (Top set of panels) spilling case. (Bottom set of panels) plunging case. Note that $t$=0 s is referenced to the time of maximum water level at a given location.

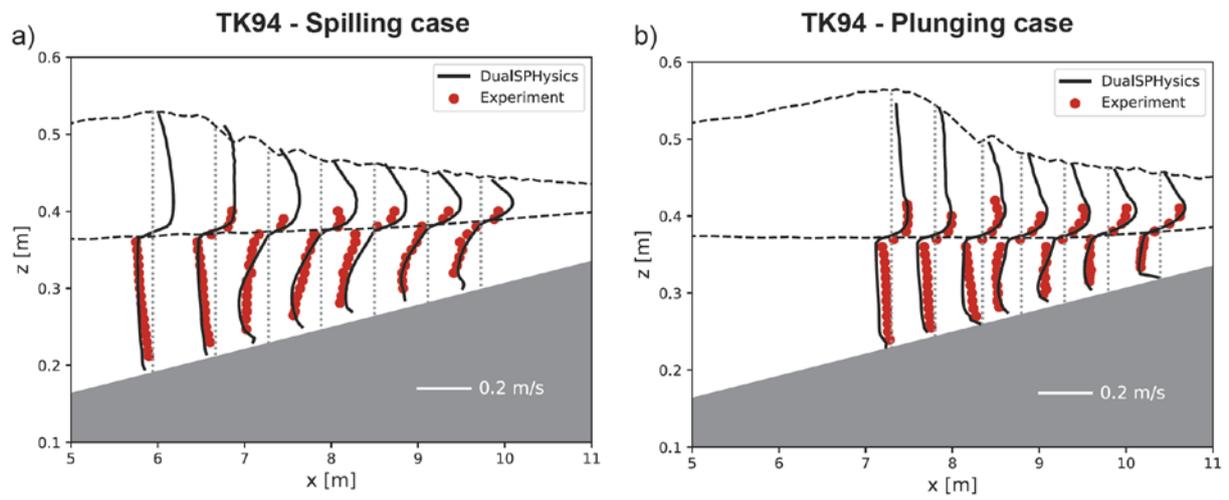

**Figure 5.** Cross-shore horizontal mean current ($u$) profiles for the a) spilling and b) plunging wave cases in TK94. The horizontal dashed lines denote the ensemble averaged crest and trough elevations. The vertical dotted lines coincide with $u=0$.

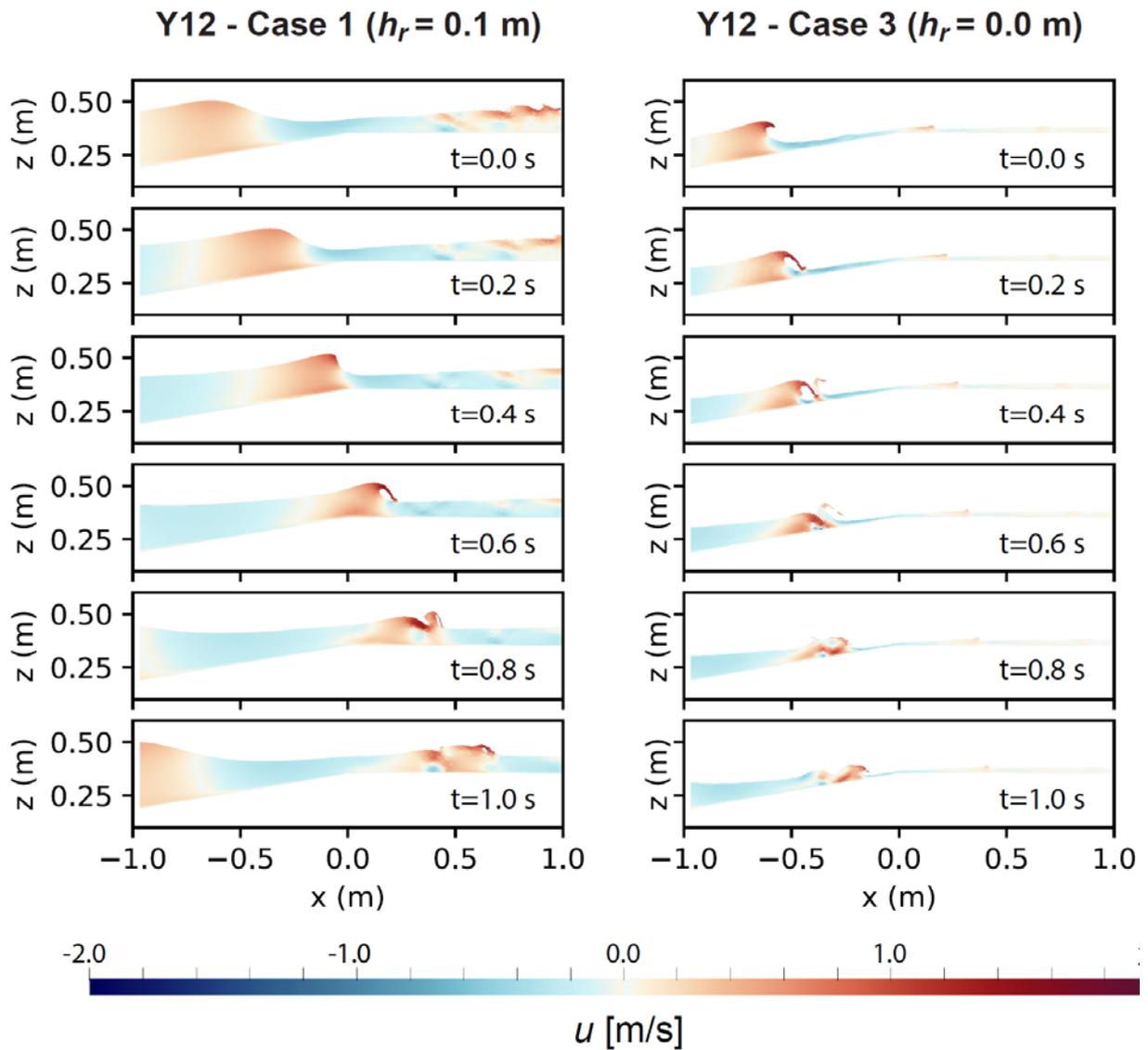

**Figure 6.** Wave breaking on a plane beach from Y12 with colours denoting the horizontal ($u$) velocity component. (Left column) Case 1 with $h_r$=0.1 m. (Right column) Case 3 with $h_r$=0.0 m. In both cases the velocity output is displayed at 0.2 s interval, where time $t$=0 s is arbitrarily assigned to the first figure in the sequence.

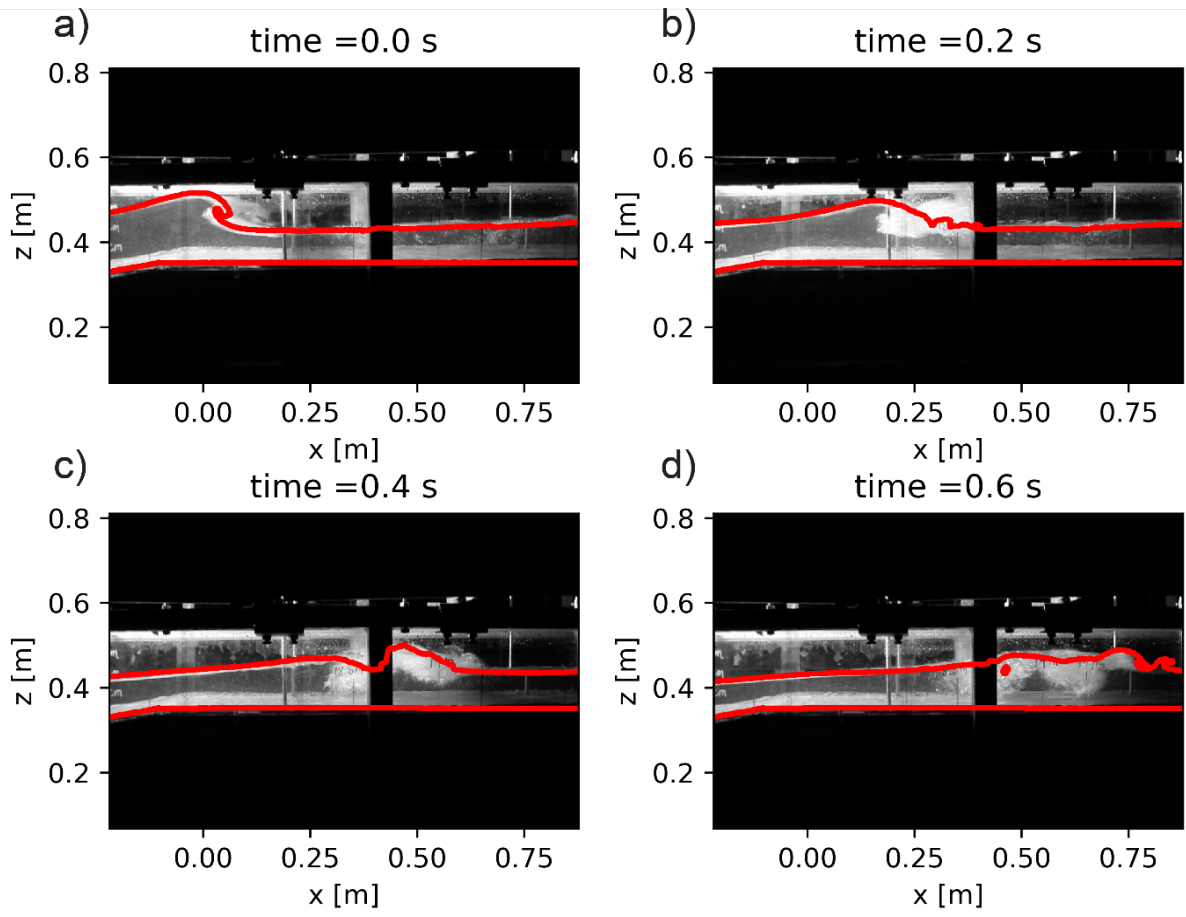

**Figure 7. Wave breaking sequence Case 1 of Y12 with the modelled free surface profile superimposed, plotted at 0.2 second intervals.**

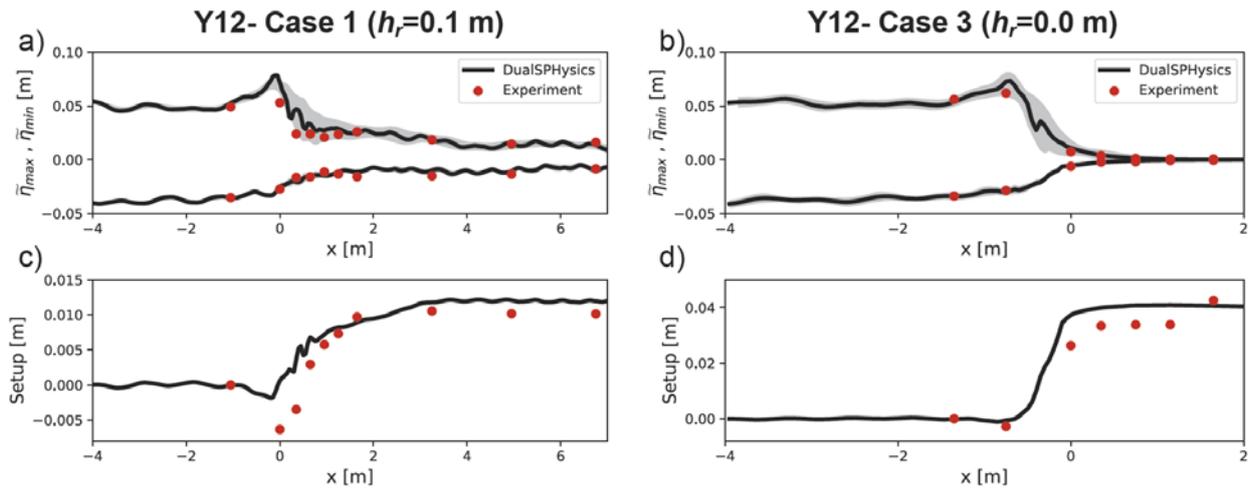

**Figure 8. Wave height envelope and wave setup evolution for the two Y12 fringing reef cases. (Left column) Case 1 with $h_r$=0.1 m. (Right column) Case 3 with $h_r$=0.0 m. (Top row) envelope of the maximum crest elevation ($\tilde{\eta}_{max}$) and minimum trough elevation ($\tilde{\eta}_{min}$) derived from the wave ensemble-averaged water levels. (Bottom row) wave setup. Grey regions denote one standard deviation of the water elevations from the ensemble (wave) average.**

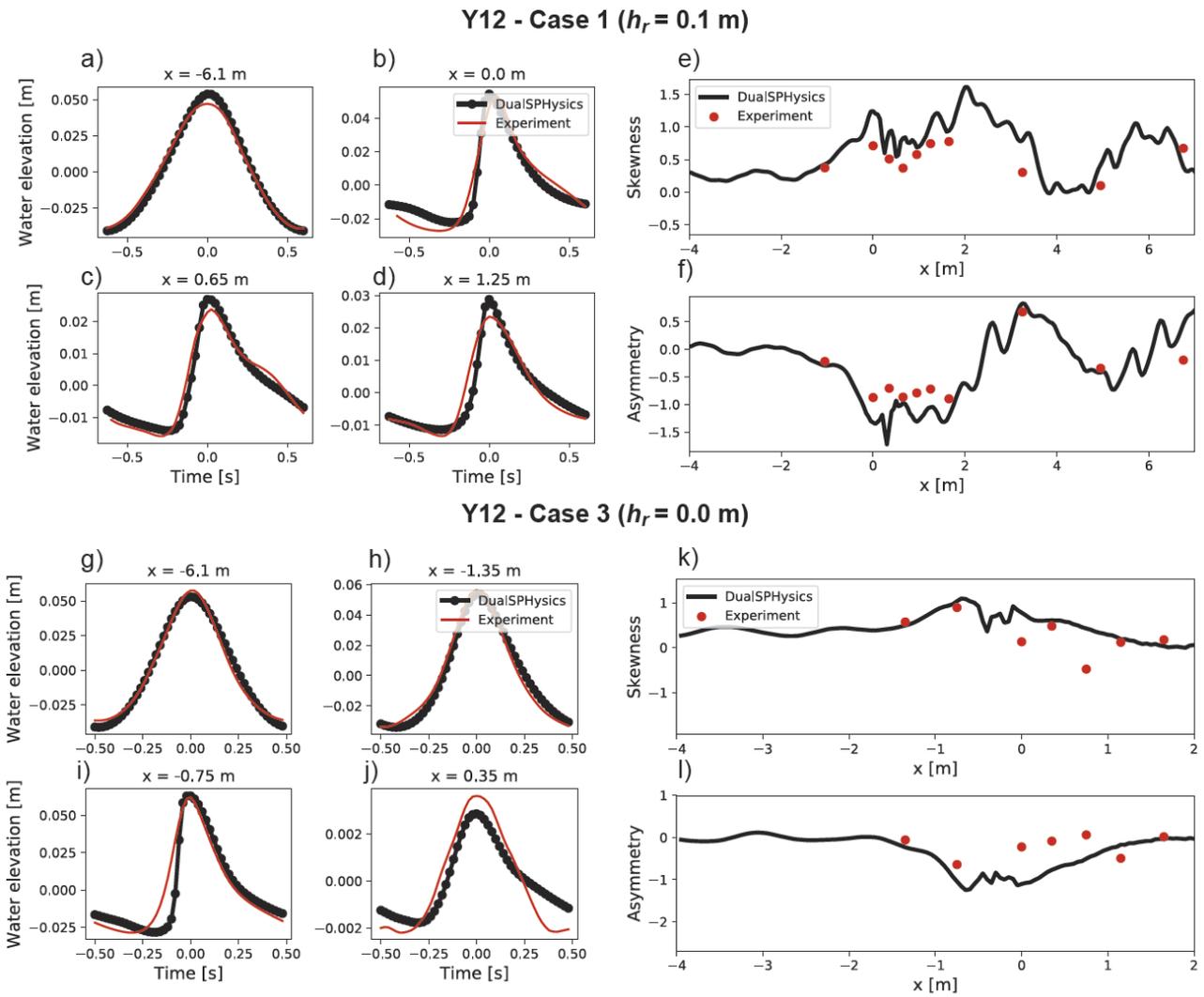

Figure 9. Water surface elevation (wave ensemble-averaged) and wave shape (skewness and asymmetry) evolution over the fringing reef for Y12. (Top set of panels) Case 1 with $h_r$=0.1 m. (Bottom set of panels) Case 3 with $h_r$=0.0 m. Note that $t$=0 s is referenced to the time of maximum water level at a given location.

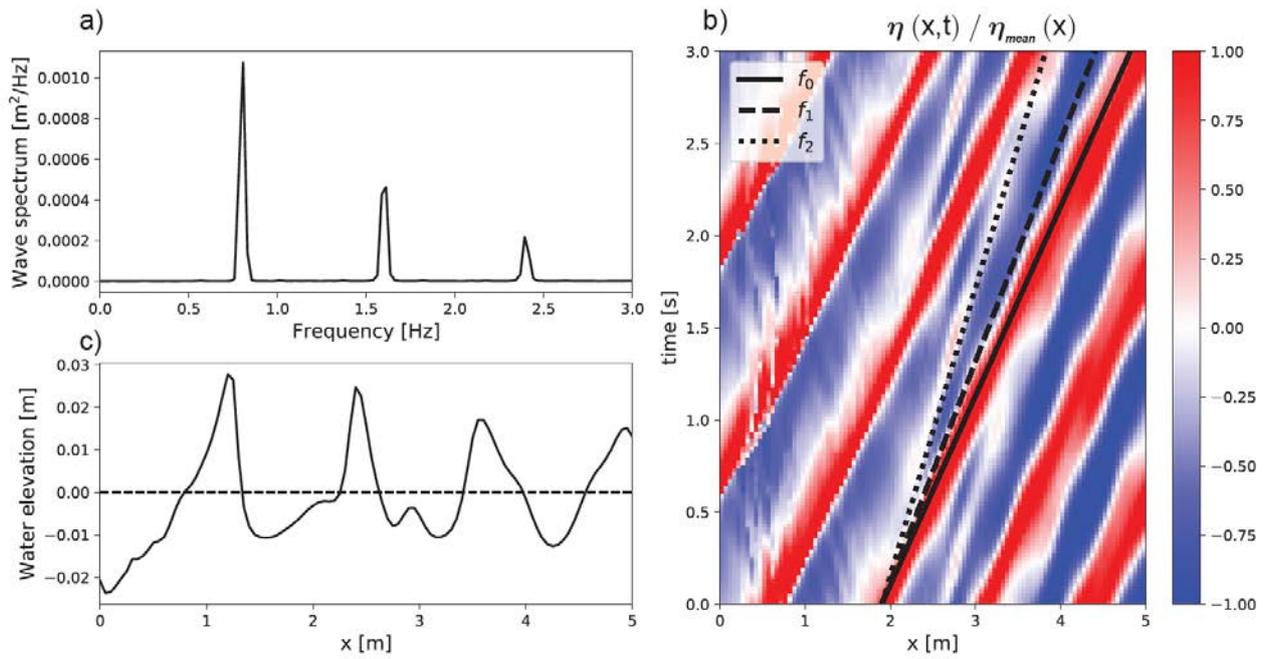

**Figure 10.** a) Wave spectrum on the reef flat ($x$=2 m). b) Timeseries of the instantaneous water levels across the reef flat over a 3 second period, normalized by the mean wave amplitude at each location $\eta_{mean} \equiv \left( \left| \tilde{\eta}_{max} \right| + \left| \tilde{\eta}_{min} \right| \right)/2$. The three curves denote the estimated wave paths based on linear wave celerity for the incident wave ($f_0$) and the first two super-harmonics ($f_1$ and $f_2$). c) Water elevation plotted across the reef based on the wave ensemble-averaged elevation $\tilde{\eta}$ (plotted at arbitrary phase) showing the change in wave shape across the reef.

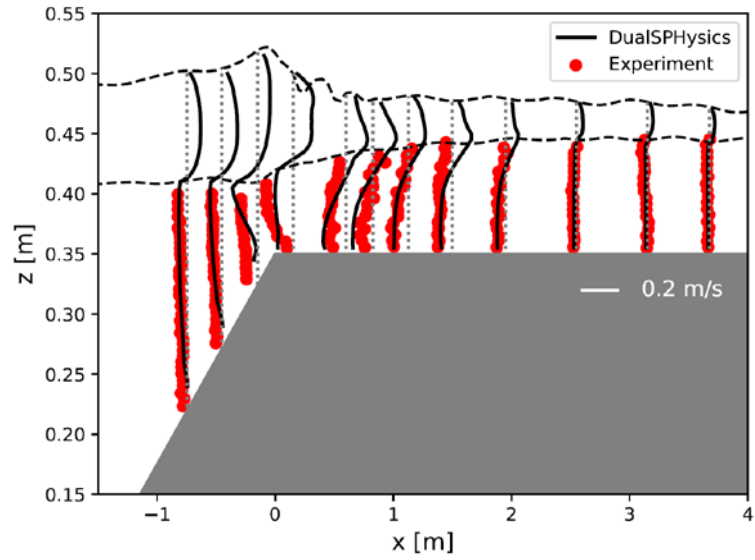

**Figure 11. Cross-shore horizontal mean current (*u*) profiles for Case 1 in Y12. The horizontal dashed lines denote the ensemble averaged crest and trough elevations. The vertical dotted lines coincide with *u*=0.**

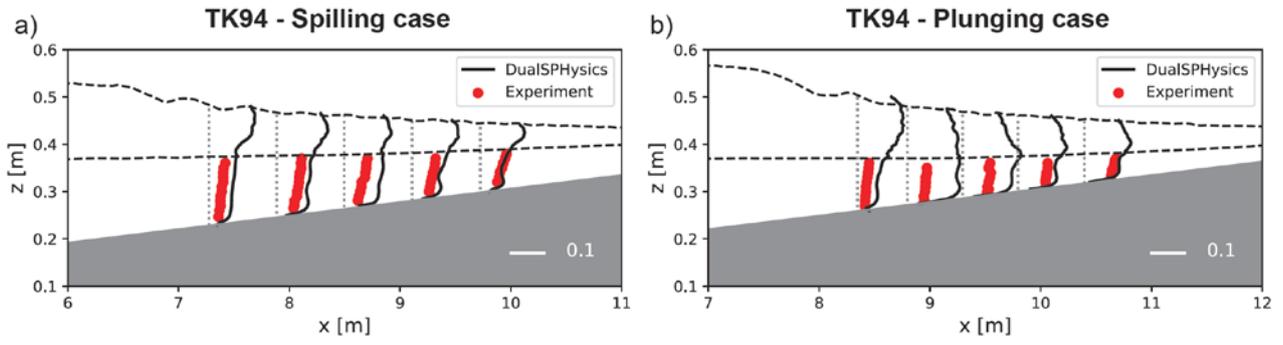

Figure 12. Normalized mean turbulent kinetic energy profiles ($\bar{k}_t^{1/2}/c$) for the a) spilling and b) plunging cases in TK94. Each vertical dotted line denotes the measurement location for comparison, where $\bar{k}_t^{1/2}/c = 0$.

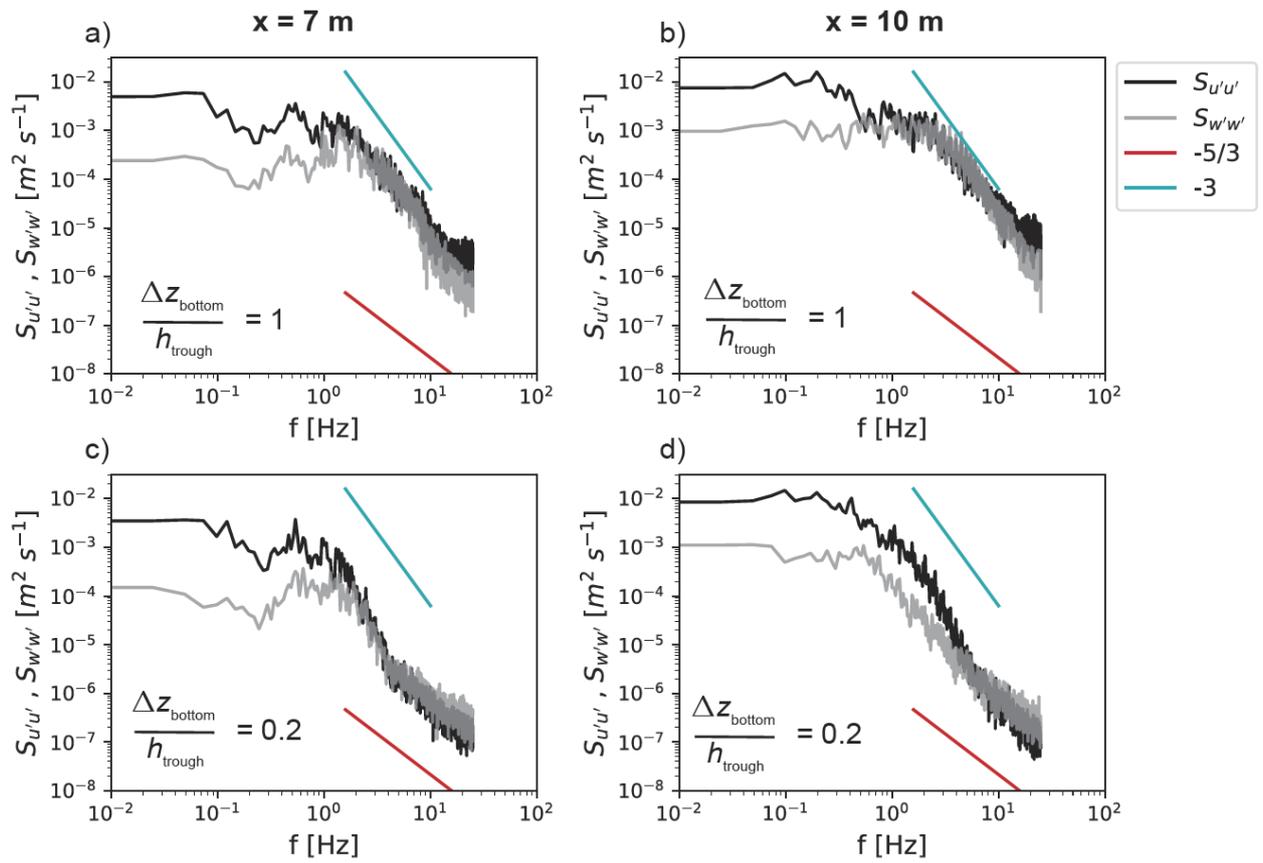

**Figure 13.** Turbulent velocity spectra for horizontal component $S_{u'u'}$ and vertical component $S_{w'w'}$ at locations within the outer surf zone (*x*=7 m; left column) and inner surf zone (*x*=10 m; right column) at two vertical locations coinciding with the trough elevation (top row) and near the bottom (bottom row). Note that the -3 slope (red) and -5/3 slope (blue) in log-log space are shown in arbitrary locations to support visual assessment of trends in the data.

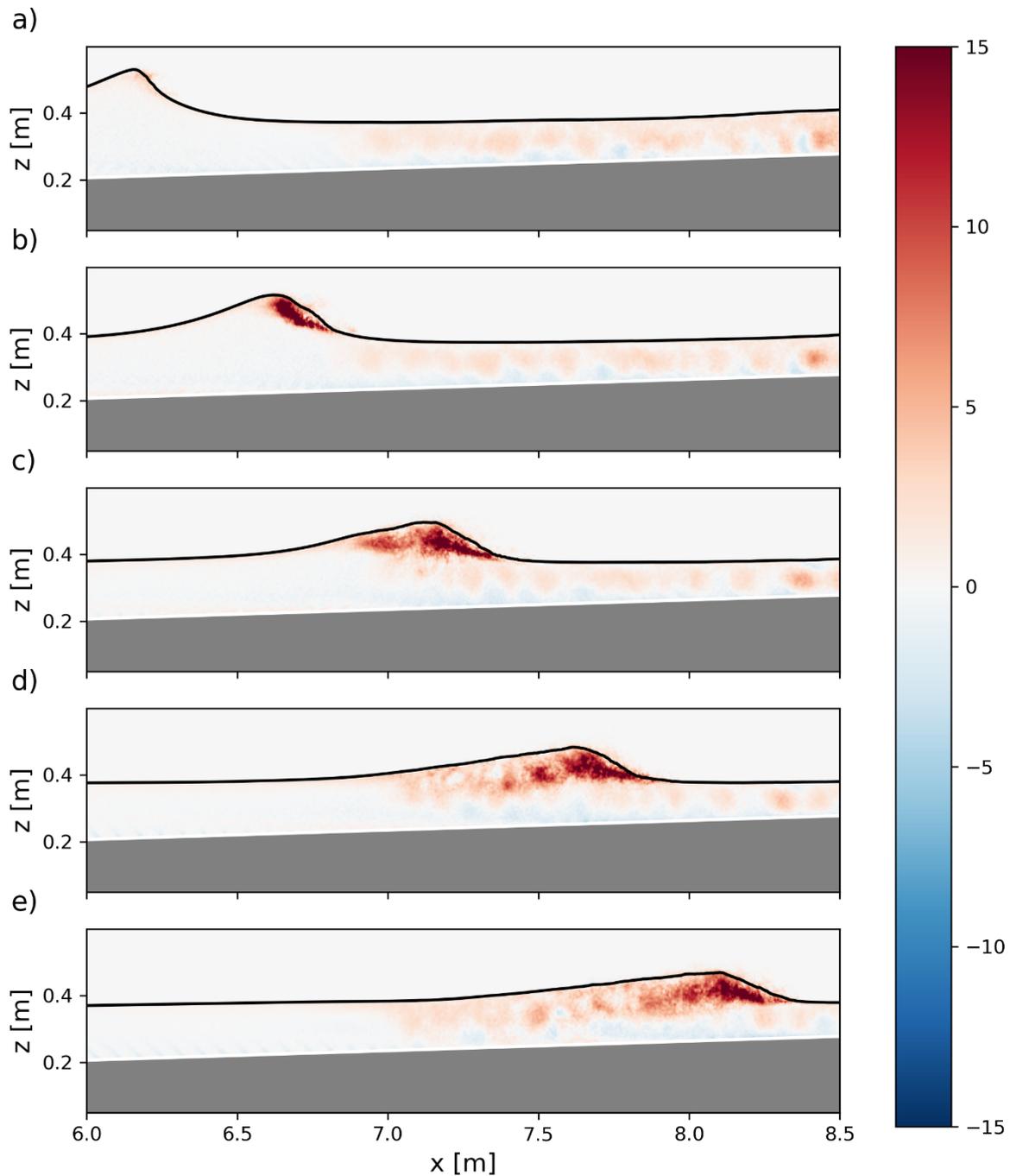

**Figure 14.** Phase-averaged vorticity [s$^{-1}$] for the spilling case of TK94, plotted at 0.3 second interval within the surf zone region. Note that colorbar values exceeding the range +/- 15 s$^{-1}$ are capped at these limiting values and a sign convention is used where positive values represent clockwise motions.

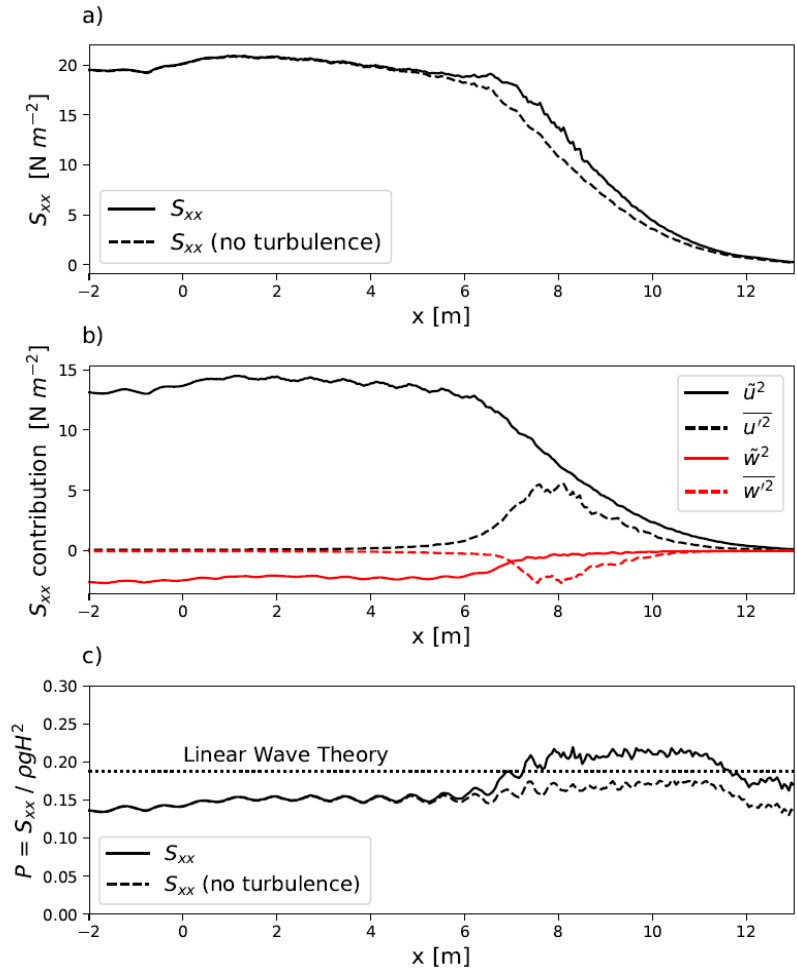

**Figure 15. a) Cross-shore distributions of radiation stresses for the spilling case in TK94 with and without turbulence contributions included. b) Individual contributions to the radiation stresses in Eq. (12). c) Ratio of radiation stresses to $\rho g H^2$ (where the ratio equals 3/16 for linear waves) for cases with and without the turbulence contribution to radiation stresses included.**

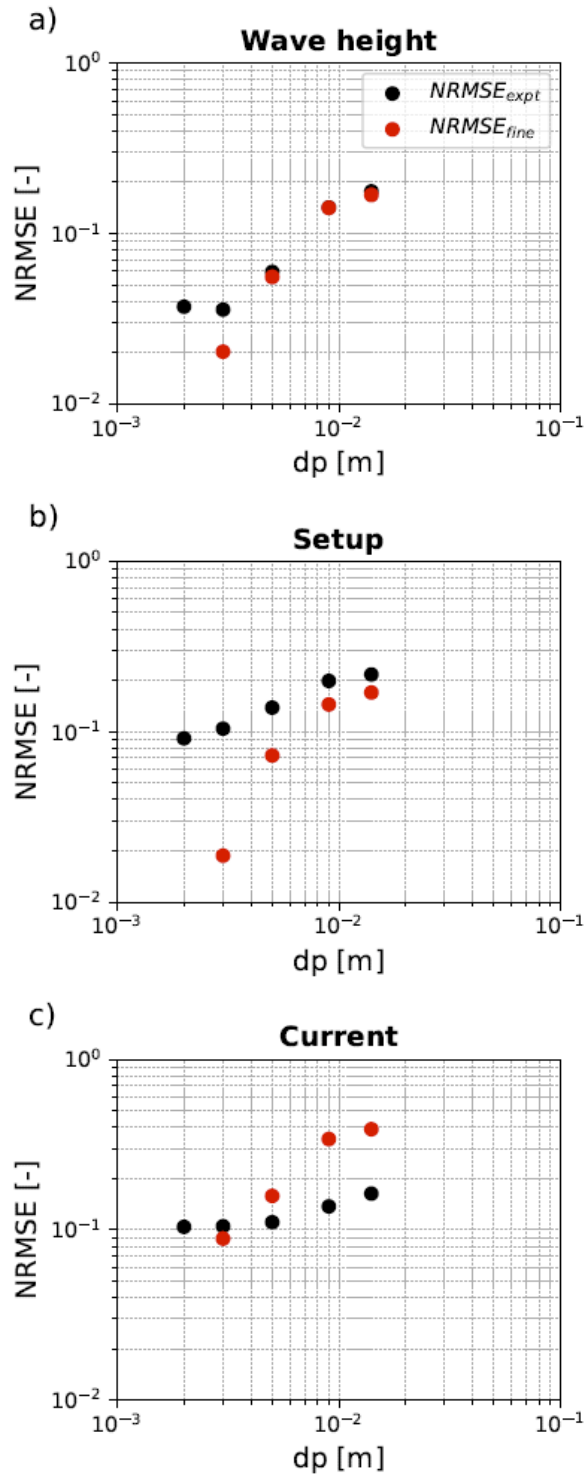

**Figure A1.** Convergence behaviour of the normalised root mean squared error (NRMSE) with varying initial inter-particle distance (*dp*) through comparison of the a) wave heights, b) setup d, and c) mean current profiles. *NRMSE$_{expt}$* (black dots) and *NRMSE$_{fine}$* (red dots) denote the NRMSE calculated using Eq. (A.2) by referencing the model predictions against the experimental data and finest resolution results (*dp*=0.002 mm), respectively.